\documentclass[twocolumn]{aastex62}

\usepackage{amsmath}
\usepackage{graphicx}
\usepackage{lipsum}
\usepackage{hyperref}
\usepackage{mathtools, nccmath}
\usepackage{longtable}

\newcommand{\dustem}{\texttt{DustEM}}

\newcommand\rev[1]{{\color{black} #1}}

\graphicspath{{./}{figures/}}

\submitjournal{AAS Journals}
\accepted{May 9, 2023}

\shorttitle{IR Imaging of Circumstellar Dust around WR~140 with Subaru and Keck}
\shortauthors{Lau et al.}

\begin{document}

\title{From Dust to Nanodust: Resolving Circumstellar Dust from the Colliding-Wind Binary Wolf-Rayet (WR) 140}

\correspondingauthor{Ryan Lau}
\email{ryan.lau@noirlab.edu}

\author{Ryan M.\ Lau}
\affil{NSF’s NOIRLab, 950 N. Cherry Ave., Tucson, AZ 85719, USA}
\affil{Institute of Space \& Astronautical Science, Japan Aerospace Exploration Agency, 3-1-1 Yoshinodai, Chuo-ku, Sagamihara, Kanagawa 252-5210, Japan}
\author{Jason Wang}
\affil{Center for Interdisciplinary Exploration and Research in Astrophysics (CIERA), Northwestern University, 1800 Sherman Ave, Evanston, IL 60208, USA}
\author{Matthew J.\ Hankins}
\affil{Department of Physical Sciences, Arkansas Tech University, 1701 N. Boulder Avenue, Russellville, AR 72801, USA}
\author{Thayne Currie}
\affil{Subaru Telescope, National Astronomical Observatory of Japan, 650 North A‘ohoku Place, Hilo, HI 96720, USA}
\author{Vincent Deo}
\affil{Subaru Telescope, National Astronomical Observatory of Japan, 650 North A‘ohoku Place, Hilo, HI 96720, USA}
\author{Izumi Endo}
\affil{Department of Astronomy, School of Science, University of Tokyo, 7-3-1 Hongo, Bunkyo-ku, Tokyo 113-0033, Japan}
\author{Olivier Guyon}
\affil{Subaru Telescope, National Astronomical Observatory of Japan, 650 North A‘ohoku Place, Hilo, HI 96720, USA}
\affil{Steward Observatory, University of Arizona, Tucson, AZ 85721, USA}
\affil{Astrobiology Center of NINS, 2-21-1 Osawa, Mitaka, Tokyo 181-8588, Japan}
\author{Yinuo Han}
\affil{Institute of Astronomy, University of Cambridge, Madingley Road, Cambridge CB3 0HA, UK}
\author{Anthony P.\ Jones}
\affil{Institut d’Astrophysique Spatiale, CNRS, Univ. Paris-Sud, Université Paris-Saclay, Bât. 121, 91405 Orsay Cedex, France}
\author{Nemanja Jovanovic}
\affil{Department of Astronomy, California Institute of Technology, Pasadena, CA 91125, USA}
\author{Julien Lozi}
\affil{Subaru Telescope, National Astronomical Observatory of Japan, 650 North A‘ohoku Place, Hilo, HI 96720, USA}
\author{Anthony F.~J.~Moffat}
\affil{Département de Physique, Université de Montréal, C.P. 6128, succ. centre-ville, Montréal (Qc) H3C 3J7, Canada; and Centre de Recherche en Astrophysique du Québec, Canada}
\author{Takashi Onaka}
\affil{Department of Physics, Faculty of Science and Engineering, Meisei University, 2-1-1 Hodokubo, Hino, Tokyo 191-8506, Japan}
\affil{Department of Astronomy, School of Science, University of Tokyo, 7-3-1 Hongo, Bunkyo-ku, Tokyo 113-0033, Japan}
\author{Garreth Ruane}
\affil{Department of Astronomy, California Institute of Technology, Pasadena, CA 91125, USA}
\author{Andreas A.\ C.\ Sander}
\affil{Zentrum für Astronomie der Universit{\"a}t Heidelberg, Astronomisches Rechen-Institut,\\ M{\"o}nchhofstr. 12-14, 69120 Heidelberg, Germany}
\author{Samaporn Tinyanont}
\affiliation{Department of Astronomy and Astrophysics, University of California, 1156 High St., Santa Cruz, CA 95064, USA}
\author{Peter G. Tuthill}
\affil{Sydney Institute of Astronomy (SIfA), School of Physics, The University of Sydney, NSW 2006, Australia}
\author{Gerd Weigelt}
\affil{Max-Planck-Institut für Radioastronomie, Auf dem Hügel 69, 53121 Bonn, Germany}
\author{Peredur M.~Williams}
\affil{Institute for Astronomy, University of Edinburgh, Royal Observatory, Edinburgh EH9 3HJ, UK}
\author{Sebastien Vievard}
\affil{Subaru Telescope, National Astronomical Observatory of Japan, 650 North A‘ohoku Place, Hilo, HI 96720, USA}
\affil{Astrobiology Center of NINS, 2-21-1, Osawa, Mitaka, Tokyo, 181-8588, Japan}


\begin{abstract}

Wolf-Rayet (WR) 140 is the archetypal periodic dust-forming colliding-wind binary that hosts a carbon-rich WR (WC) star and an O-star companion with an orbital period of 7.93 years and an orbital eccentricity of 0.9. Throughout the past several decades, multiple dust-formation episodes from WR~140 have been observed that are linked to the binary orbit and occur near the time of periastron passage. Given its predictable dust-formation episodes, WR~140 presents an ideal astrophysical laboratory for investigating the formation and evolution of dust in the hostile environment around a massive binary system. In this paper, we present near- and mid-infrared (IR) spectroscopic and imaging observations of WR~140 with Subaru/SCExAO+CHARIS, Keck/NIRC2+PyWFS, and Subaru/COMICS taken between 2020 June and Sept that resolve the circumstellar dust emission linked to its most recent dust-formation episode in 2016 Dec. 
Our spectral energy distribution (SED) analysis of WR~140's resolved circumstellar dust emission reveals the presence of a hot ($T_\mathrm{d}\sim1000$ K) near-IR dust component that is co-spatial with the previously known and cooler ($T_\mathrm{d}\sim500$ K) mid-IR dust component composed of $300\text{--}500$ \AA-sized dust grains. We attribute the hot near-IR dust emission to the presence of nano-sized (“nanodust”) grains and suggest they were formed from grain-grain collisions or the rotational disruption of the larger grain size \rev{population} by radiative torques in the strong radiation field from the central binary. Lastly, we speculate on the astrophysical implications of nanodust formation around colliding-wind WC binaries, which may present an early source of carbonaceous nanodust in the interstellar medium.

\end{abstract}


\section{Introduction}
\label{sec:intro}

Dust grains are a ubiquitous component in the Universe and are present even in high-redshift galaxies less than 1~Gyr after the Big Bang (\citealt{Lesniewska2019,Hashimoto2019,Bakx2020}). The infrared (IR) emission from dust also presents a valuable tracer of star \rev{and planet} formation and the chemical enrichment of the interstellar medium (ISM) of galaxies across cosmic time.
However, the origin(s) of substantial quantities of dust measured in the ISM of both local and high-redshift galaxies remain(s) uncertain (e.g.,~\citealt{Dwek2011}).
Although the leading sources of dust are thought to be core-collapse supernovae (SNe; e.g.,~\citealt{Dwek1998,Gall2011,Lesniewska2019}) or Asymptotic Giant Branch (AGB; e.g.,~\citealt{Srinivasan2009,Valiante2009,Boyer2012}) stars, dust formation from Wolf-Rayet (WR) stars presents an early and potentially significant source of dust \citep{Marchenko2017,Lau2020a}.

Carbon-rich WR (WC) stars are capable of producing copious amounts of carbon-rich dust ($\dot{M}_d\sim10^{-10}-10^{-6}$ M$_\odot$ yr$^{-1}$; \citealt{Zubko1998,Lau2020a}) and form on relatively short \rev{evolutionary} timescales ($\sim$Myr; \citealt{Crowther2007}).
Recent results from \citet{Franco2021} have also highlighted WR stars as potentially important sources of the chemical enrichment of galaxies in the early Universe. 
\rev{However, understanding} dust formation around WC stars has been challenging given their hostile circumstellar environment from their fast winds and strong radiation field. \rev{Notably,} the condensation of carbon grains requires much higher densities (greater than a factor of $10^3$) than what is expected in the isotropic winds of a single WC star (e.g.,~\citealt{Hackwell1979,Cherchneff2000}).

Observational and theoretical studies have demonstrated that dust formation in such environments can be enabled by colliding winds between the WC star and a massive companion such as an O-type star \citep{Usov1991,Williams2009}.
The \rev{WC7pd +O5.5fc} binary system WR~140 (= HD~193793), which briefly forms dust around periastron passage, has provided one of the most important astrophysical laboratories for investigating colliding-wind dust formation given its well-defined $P_\mathrm{orb}=7.93$ yr highly eccentric ($e=0.90$) orbit and the wealth of multi-wavelength observations taken over the past several decades \citep{Williams1978,Williams1990,Monnier2011,Sugawara2015,Thomas2021,Williams2021,Pollock2021}. For this reason, WR~140 has been dubbed as the ``Rosetta Stone" of colliding-wind systems \citep{Tuthill2003}.

A comprehensive analysis on the orbitally modulated dust formation by WR~140 was conducted by \citet{Williams2009}, who presented multi-year IR imaging observations that captured the evolving morphology of the rapidly expanding circumstellar dust after its 2001 periastron passage dust-formation episode. 
With spatially resolved IR images of WR~140, \citet{Williams2009} presented several possible models of the circumstellar dust morphology and conducted a single-component isothermal dust SED analysis that indicated the presence of $\sim0.01$ $\mu$m-sized carbonaceous grains.
Interestingly, they also observed a decrease in the circumstellar dust mass by a factor of $\gtrsim3$ over several years, which suggests dust destruction in WR~140's circumstellar environment.
\rev{Recent observations of WR~140 with the James Webb Space Telescope (JWST; \citealt{Gardner2006}) by \citet{Lau2022} revealed over 17 nested circumstellar dust shells extending beyond $\sim70,000$ au from the central binary demonstrating the survival of the dust grains over at least 17 periods ($\gtrsim130$ yr). Notably, JWST detected broad mid-IR emission features at 6.4 and 7.7 $\mu$m from spatially resolved circumstellar dust around WR~140 indicating a composition of carbon-rich aromatic compounds. }
Given the revisited significance of WC binaries as prominent sources of dust \citep{Lau2020a} \rev{and the latest results from JWST \citep{Lau2022}, an investigation of the evolution and physical properties of the circumstellar dust around WR~140 is well motivated.} 

New significant technological developments in ground-based extreme adaptive optics (AO) imaging tuned to obscured, red astrophysical sources have opened a new window for investigations of dust-production from WC binaries like WR~140. 
For example, the Keck Observatory has recently deployed an IR pyramid wavefront sensor (PyWFS) on the Keck II AO system and the vortex coronagraph on the NIRC2 instrument \citep{Bond2020}. 
The Subaru Coronagraphic Extreme Adaptive Optics Project (SCExAO, \citealt{Jovanovic2015}) on the Subaru Telescope also presents a cutting-edge extreme AO platform with a Pyramid wavefront sensor that operates at red optical wavelengths. 
The AO correction combined with the high contrast achievable with the new technologies on Keck and Subaru enable new studies of dust production and mass-loss from dust-forming WC binaries.

In this paper, we report the following near-contempor-aneous high-angular-resolution IR observations of WR~140: spectroscopic imaging between 1.2 \rev{and} 2.4 $\mu$m with the Coronagraphic High Angular Resolution Imaging Spectrograph (CHARIS, \citealt{Groff2016}) behind the SCExAO system, $L'$-band imaging with Keck/NIRC2+PyWFS, and $N$-band imaging with the Cooled Mid-Infrared Camera and Spectrograph (COMICS) on the Subaru Telescope.
Our observations resolve the circumstellar dust around WR~140 and present some of the first extreme-AO observations of the circumstellar environment around massive stars.

The paper is organized as follows. First, in Sec.~\ref{sec:Obs} we present and describe the WR~140 observations and data reduction with Subaru/SCExAO/CHARIS, Keck/NIRC2+PyWFS, and Subaru/COMICS. In Sec.~\ref{sec:RA}, we analyze the morphology of the IR emission from circumstellar dust around WR~140 and conduct a dust SED model analysis to investigate the dust properties. The dust SED analysis reveals the presence of very small nano-sized (``nanodust") grains around WR~140. Lastly, in Sec.~\ref{sec:disc}, we discuss the possible mechanisms that could lead to produce nanodust around WR~140 and the astrophysical implications of nanodust formation around WC binaries.

\begin{figure*}[t!]
    \includegraphics[width=.96\linewidth]{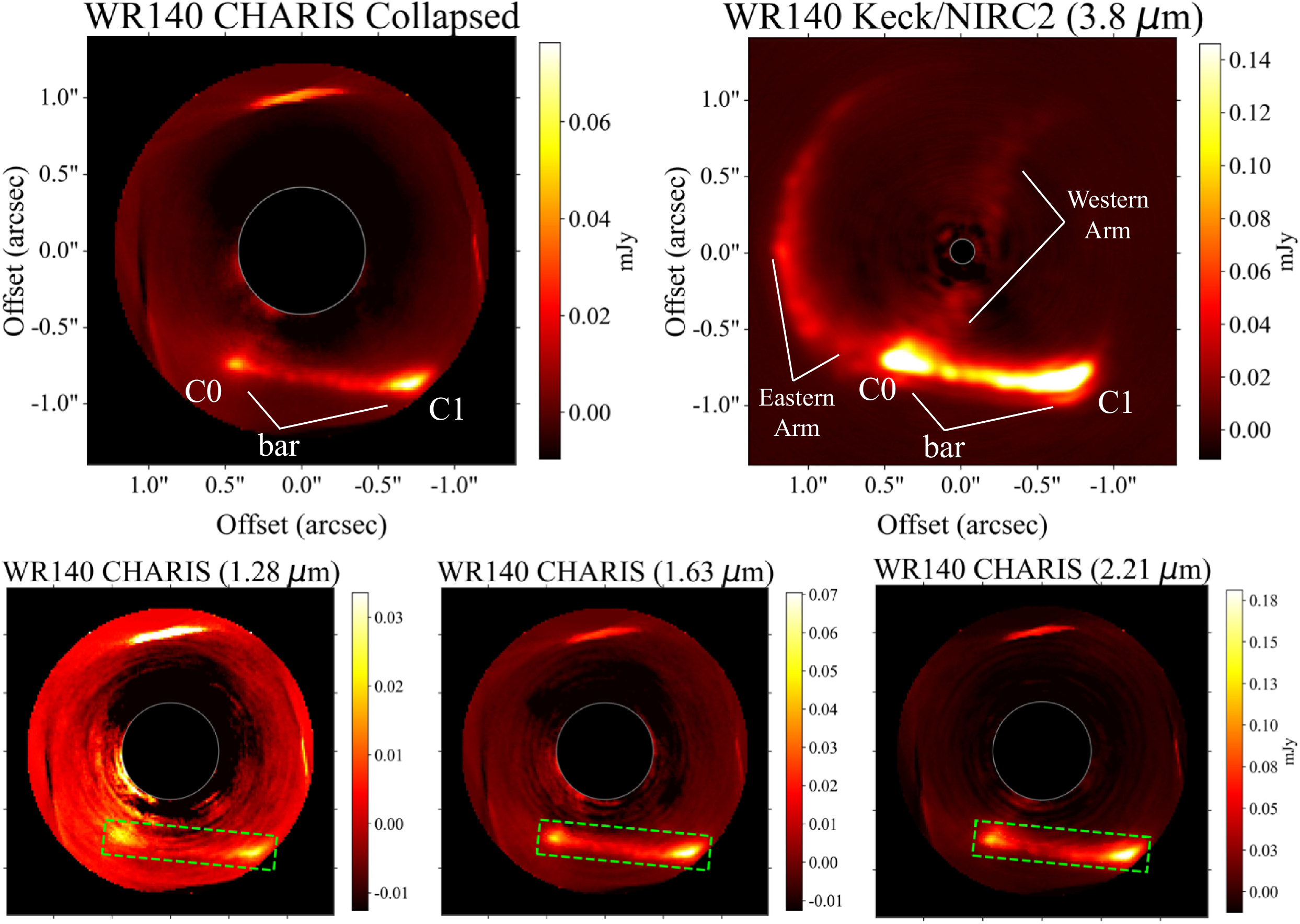}
    \caption{ (\textit{Top Left}) Subaru/SCExAO/CHARIS collapsed  1.2--2.4 $\mu$m image and (\textit{Top Right}) Keck/NIRC2 $L'$-band image of WR~140 taken on 2020 Sept 2 and 2020 Aug 18, respectively. The central masks in the images correspond to the $3\times$ the 113 mas inner working angle of the CHARIS observations and the 80 mas inner working angle of the NIRC2 observations.  The positions of persistent dust features identified by \citet{Williams2009} as well as the Western Arm are overlaid on each image. The bright and linear horizontal feature at a vertical offset of +1\farcs0 is likely an artifact from the data reduction. \rev{The absence of this horizontal feature in the NIRC2 observations suggests it is indeed an artifact.} (\textit{Bottom Row}) CHARIS images of WR~140 at individual slices corresponding to 1.28, 1.63\rev{,} and 2.21 $\mu$m with a spectral resolution of $R\sim19$. The rectangular aperture used to extract the flux density of the `bar' is overlaid on the CHARIS image slices. }
    \label{fig:Collage}
\end{figure*}

\begin{deluxetable*}{p{4.6cm}p{2.5cm}p{3.3cm}p{2.6cm}}
\tablecaption{WR~140 Observation Summary}
\tablewidth{0pt}
\tablehead{ & Subaru/CHARIS & Keck/NIRC2+PyWFS & Subaru/COMICS}\
\startdata
Observing Mode & Broadband+Lyot 113 mas & $L'$ Vortex Coronagraph & Imaging \\
Observation Date (UTC) & 2020 Sept 2 \newline (MJD 59094.2) & 2020 Aug 18 \newline (MJD 59079.4)  & 2020 June 27 \newline (MJD 59027.5) \\
WR~140 \rev{Orbital} Phase ($\varphi$) & 0.467 & 0.462 & 0.444 \\
Wavelength ($\mu$m) & 1.15--2.39 & 3.8 & 11.7\\
Spec.~Resolving Power~($\lambda/\Delta \lambda$) & $\sim 19$ & $\sim5$ & $\sim12$\\
Angular Res.~(FWHM) & 0.03--0.06\arcsec & 0.074\arcsec & 0.35\arcsec\\
Pixel Scale (mas) & \rev{16.4} & 9.94 & 130\\
Field of View & $\sim$2\arcsec$\times$ 2\arcsec & 10\arcsec$\times$10\arcsec & 42\arcsec$\times$ 32\arcsec\\
\enddata
\tablecomments{Summary of WR~140 observations with Subaru/SCExAO/CHARIS, Keck/NIRC2, and Subaru/COMICS. The WR~140 orbital phase ($\varphi$) is derived from the ephemeris derived by \citet{Monnier2011} with the \citet{Fahed2011} prior.}
\label{tab:Obs}
\end{deluxetable*}

\section{Observations and Data Reduction}
\label{sec:Obs}

\subsection{Near-IR Integral Field Spectroscopy with Subaru SCExAO/CHARIS}
We observed WR 140 (J2000 R.A. 20:20:27.98, Dec. +43:51:16.29; \citealt{Lindegren2021}) on 2020 September 2 UT with SCExAO coupled to the CHARIS integral field spectrograph operating in low-resolution/broadband mode, covering the major $JHK$ passbands simultaneously ($\lambda_{\rm o}$ = 1.16--2.39 $\mu$m) and using the $\sim$0\farcs{}113 radius Lyot coronagraphic mask.   Seeing values measured from the Mauna Kea Weather Center ranged between 0\farcs{}5 and 0\farcs{}8 but with high (20--25 mph) winds.   The observations covered $-$2.3 \rev{to} $-$0.94 in hour angle and were conducted in ``pupil tracking"/angular differential imaging mode \citep{Marois2006} to allow the sky to rotate on the CHARIS detector with time ($\Delta$PA = 30$^{o}$).    As WR 140 was slightly off center from the CHARIS detector, we have full 360$^{o}$ coverage out to $\sim$0\farcs{}96 instead of the more typical 1\farcs{}05.   For astrometric calibration, we took four exposures with the artificial satellite spots \citep{Jovanovic2015b} turned on and obtained the rest of our observations without satellite spots.   We observed the K0-star HD 194479 afterwards as a PSF reference star and for spectrophotometric calibration.

We used the standard CHARIS pipeline \citep{Brandt2017} to extract CHARIS data cubes from the raw data.  For subsequent data reduction, we used the CHARIS Data Processing Pipeline \citep{Currie2020,Currie2020Sci}.   For spectrophotometric calibration drawn from \rev{the} HD 194479 data, we adopted a Kurucz stellar atmosphere model appropriate for a K0V star.   Simple inspection showed that the `bar' \rev{and C1} emission \rev{features are} bright (Fig.~\ref{fig:Collage}, \textit{Top Left}), especially at longer wavelengths, and detected with conservative PSF subtraction methods.  Thus, to subtract the PSF, we used a full-frame implementation of reference star differential imaging (RDI) using the \textit{Karhunen-Loève Image Projection} \citep[KLIP;][]{Soummer2012} algorithm as in \citet{Currie2019} retaining only the first 3 KL modes.  Separately, we performed a simple classical spectral differential imaging (SDI) reduction of the post-RDI/KLIP residuals.   While classical SDI produced some attenuation of the bar at the shortest wavelengths, it fully preserved the overall shape and morphology of the emission feature with reduced residual speckles.  
The bright and linear horizontal emission feature at a vertical offset of +1\farcs0 near the edge of the field of view is an artifact from the data reduction (Fig.~\ref{fig:Collage}, \textit{Top Left}).

To estimate flux loss in RDI/KLIP-reduced data due to oversubtraction \citep{Pueyo2016}, we inserted into our RDI/KLIP sequence the wavelength-collapsed final image produced using classical SDI on the post-RDI/KLIP residual images.   We then forward-modeled this image through our RDI/KLIP sequence to estimate signal loss.   Forward-modeling showed that the flux from the bar is well preserved, with a throughput varying from 0.90 to 0.98 across the entire bar morphology and over the full range of CHARIS channels. We therefore conservatively adopt a $1\sigma$ photometric uncertainty of $10\%$ across the CHARIS channels.

\rev{Flux densities of the C1 and the bar were extracted from the CHARIS data cubes using a circular aperture with a 250 mas radius and a rectangular aperture with a size of $1480 \times 300$ mas, respectively. The rectangular aperture for the bar was tilted by 1.7$^\circ$ south of west to match the angle of the bar feature.}

\subsection{$L$'-band Imaging with the Keck/NIRC2 Vortex Coronagraph}

Keck/NIRC2 L'-band ($\lambda_c = 3.8$ $\mu$m) imaging observations of WR~140 were obtained on 2020 August 18 UT in the program C258 (PI - M.~Hankins). The vector vortex coronagraph \citep{Vargas2016,Serabyn2017} was used to suppress the diffraction pattern of the star with an inner working angle of $\lambda/D\approx80$ mas at L'-band. The infrared pyramid wavefront sensor \citep{Bond2020} was used to obtain good adaptive optics correction as the star is much brighter in the infrared than at visible wavelengths due to extinction. Observations were taken using the narrow camera with a 10\arcsec$\times$10\arcsec~field of view and a pixel scale of 9.94 mas. QACITS, the standard focal plane wavefront sensing technique for the vector vortex coronagraph, was used to obtain observations of the system while keeping the star aligned behind the coronagraph \citep{Huby2017}. In total, we obtained 58 coronagraphic images, each consisting of 100 coadds of 0.18 s exposures. As standard in the QACITS sequence, we also obtained sky background images and unocculted images of the star for calibration. In addition to WR~140, we used the same configuration to obtain 15 frames on two reference stars: the K5-star HD 181681 immediately before, and the K0-star HD 194479 immediately afterwards. 

We used the vortex pre-processing pipeline \citep{Xuan2018} to perform bad-pixel correction, flat fielding, thermal background subtraction, and co-registration of the images. The reference images from HD 181681 and HD 194479 were used to subtract off the stellar diffraction pattern of WR~140 using the KLIP-RDI implementation in the open-source \texttt{pyKLIP} package \citep{Wang2015}. Given the brightness of the dust in the data, we used 1 KL mode to model the stellar diffraction pattern. The fully reduced image is shown in Figure \ref{fig:Collage} (\textit{Top Right}). To accurately measure the flux of the dust, we forward-modeled the over-subtraction caused by KLIP-RDI by projecting the KL mode onto a model of the dust, and measured a small (1--2\%) flux loss due to KLIP-RDI that we compensate for. We conservatively adopt a $1\sigma$ $L'$-band photometric uncertainty of $10\%$. 

\rev{Flux densities of the C1 and the bar were extracted from the NIRC2 image using a circular aperture with a 250 mas radius and a rectangular aperture with a size of $1490 \times 300$ mas, respectively. The rectangular aperture for the bar was tilted by 1.7$^\circ$ south of west to match the angle of the bar feature. The flux density of the C1 ``Fossil'' dust feature (see Sec.~\ref{sec:fossil}) was extracted using an elliptical aperture with a semi-major axis and semi-minor axis of 600 and 150 mas, respectively. The elliptical C1 Fossil aperture was rotated by 11.6$^\circ$ north of west to match the orientation of the dust feature.}

\subsection{Mid-IR Imaging with Subaru/COMICS}

Mid-IR imaging observations of WR~140 were obtained using the Cooled Mid-Infrared Camera and Spectrometer (COMICS; \citealt{Kataza2000,Okamoto2003}) on the Cassegrain focus of the Subaru Telescope with the N11.7 filter ($\lambda_c = 11.7$ $\mu$m, $\Delta\lambda=1.0$ $\mu$m) on 2020 June 27 UT  (Fig.~\ref{fig:COMICS}).
Individual 100 s exposures were performed using a perpendicular chopping and nodding pattern with a chop throw and nod amplitude of $10''$. The total integration time on WR~140 was 400 s. The pixel scale of the detector is $0.13''$ pixel$^{-1}$. 

Observations of the K3II-star Gamma Aql were used as photometric and PSF references based on the list of mid-IR standards in
\cite{Cohen1999}.
The measured full-width at half maximum (FWHM) in the N11.7 imaging observations of Gamma Aql was $\sim0.35''$, which is consistent with near diffraction-limited performance at 11.7 $\mu$m ($0.29''$).

The data reduction was carried out using the COMICS Reduction Software\footnote{\url{https://subarutelescope.org/Observing/DataReduction/Cookbooks/COMICS_CookBook2p2E.pdf}} and \rev{the Image Reduction and Analysis Facility (IRAF; \citealt{Tody1986,Tody1993})}. The chopping and nodding pair subtraction was employed to cancel out the background radiation and to reduce a residual pattern from the chop subtraction. Flat fielding was achieved using self-sky-flat made from each image. A conversion factor for flux density calibration was calculated by the reduced images of the standard star and its photometric data by \cite{Cohen1999}. We adopt a $1\sigma$ $N$-band photometric calibration uncertainty of $\sim10\%$.

\rev{Flux densities of the C1 and the bar were extracted from the COMICS image using a circular aperture with a 520 mas radius and a rectangular aperture with a size of $1820 \times 650$ mas, respectively. The rectangular aperture for the bar was tilted by 1.7$^\circ$ south of west to match the angle of the bar feature. The flux density of the C1 Fossil dust feature was extracted using an elliptical aperture with a semi-major axis and semi-minor axis of 1170 and 520 mas, respectively. Background ``sky'' annuli were used to measure and subtract the mid-IR background emission. A circular annulus with inner and outer radii of 780 mas and 1560 mas were used for the background annulus of C1. A rectangular annulus with a 650 mas width and the same inner dimensions and orientation as the rectangular aperture for the bar was used for the bar. Lastly, the C1 Fossil background annulus had inner radii consistent with the dimensions of the elliptical C1 Fossil aperture and width of 650 mas.}

\section{Results and Analysis}
\label{sec:RA}

\subsection{WR~140 Dust Morphology from December 2016 Periastron Passage}

\begin{figure}[t!]
    \centerline{\includegraphics[width=0.99\linewidth]{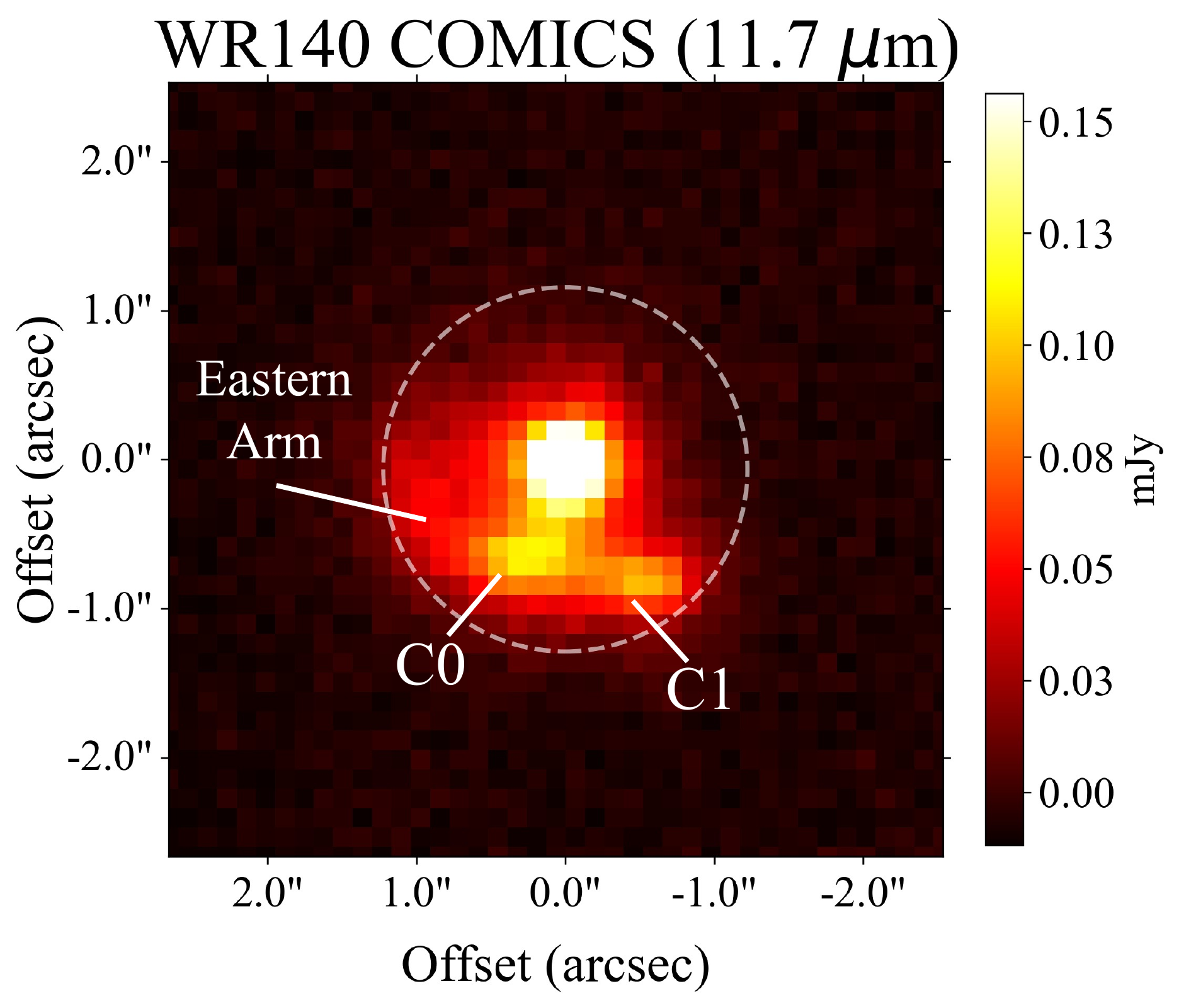}}
    \caption{Subaru/COMICS 11.7 $\mu$m imaging observations of WR~140 taken on 2020 June 27. The positions of persistent dust features are overlaid on the image. The overlaid dashed circle indicates the approximate CHARIS field-of-view.}
    \label{fig:COMICS}
\end{figure}

The Subaru/CHARIS 1.1--2.4 $\mu$m, Keck/NIRC2 3.8 $\mu$m $L'$-band, and 11.7 $\mu$m $N$-band Subaru/COMICS observations of WR~140 resolve and detect emission from dust formed during the periastron passage that occurred \rev{in} December 2016 \citep{Williams2009,Thomas2021}. These features are consistent with the ``persistent dust features'' identified by IR imaging of WR~140 in \citet{Williams2009} after its 2001 periastron passage. The most prominent features are the two bright emission peaks located 0\farcs86 and 1\farcs15 southeast and southwest of the central binary, respectively (Fig.~\ref{fig:Collage}\,\&\,\ref{fig:COMICS}).
These southeast and southwest emission peaks were labeled as `C0' and `C1,' respectively, by \citet{Williams2009}. At wavelengths longer than $\sim1.6$ $\mu$m, we detect the aforementioned faint, linear `bar' of emission between C0 and C1. The near-IR CHARIS and 3.8 $\mu$m NIRC2 observations of the bar reveal a ``clumpy'' morphology that appear to bridge C0 and C1 (Fig.~\ref{fig:Collage}). 

To the east of the bar, the NIRC2 and 11.7 $\mu$m COMICS images show a faint arc of emission that curves to the north. \citet{Williams2009} refer to this feature as the `arm,' and, more recently, \citet{Han2022} label it as the ``Eastern Arm.'' In this work, we refer to this feature as the Eastern Arm in order to distinguish it from a fainter ``Western Arm'' detected in the NIRC2 image (Fig.~\ref{fig:Collage}; \textit{Top Right}). The Eastern Arm exhibits an integrated flux density $\sim25$\% of the bar, and in the COMICS image the Eastern Arm is $\sim35$\% \rev{the intensity} of the bar. The high spatial resolution of the NIRC2 image reveals the clumpy morphology of the Eastern Arm similar to that of the bar. 
The CHARIS observations did not detect significant emission from the Eastern Arm; however, this feature is located near/slightly beyond the edge of the CHARIS field of view.

Among the three observations, only the NIRC2 image reveals the Western Arm, which extends from C0 through the central source and continues north/north-west for $\sim1$\arcsec (Fig.~\ref{fig:Collage}; \textit{Top Right}). Northward of the central source, the curvature of the Western Arm resembles that of the Eastern arm. The integrated flux of the Western Arm in the NIRC2 image is a few percent of that of the bar. Non-detection of the Western Arm in the CHARIS and COMICS observations is likely due to its faintness as well as the lower SNR towards the central source for the CHARIS observations.

\begin{figure*}[t!]
    \centerline{\includegraphics[width=0.96\linewidth]{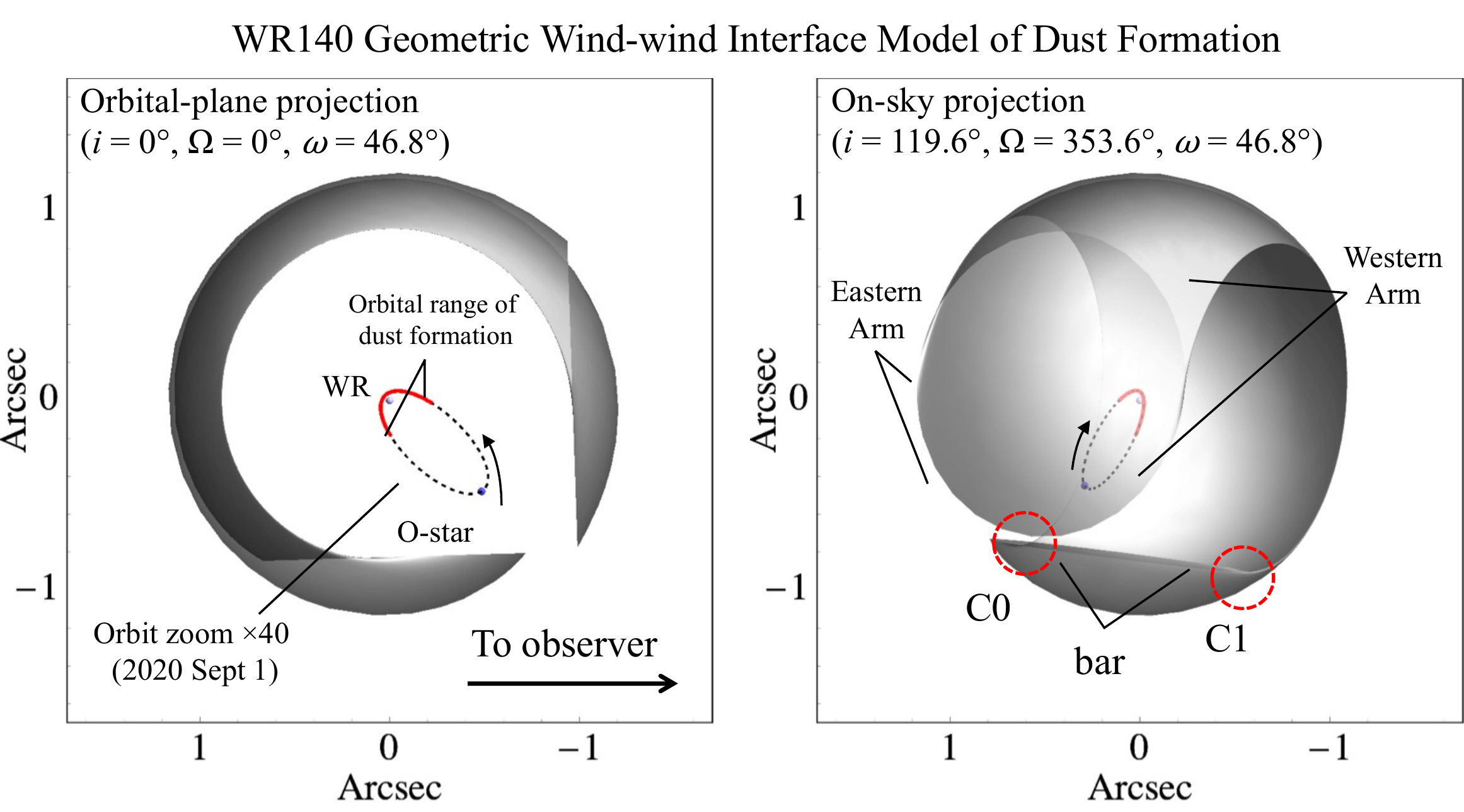}}
    \caption{(\textit{Left}) \rev{Orbital-plane} and (\textit{Right}) \rev{on-sky projections} of the geometric wind-wind interface model where dust forms in the colliding winds in the WR~140 binary around periastron passage. The positions of the dust features are overlaid on the model, where the Eastern and Western Arms are on the near side closest to the observer.
    The orbital parameters of the models are indicated above, and the parameters in the observed view are adopted from the \citet{Monnier2011} results with the \citet{Fahed2011} prior. The half-opening angle of the colliding-wind interface is assumed to be $\theta\approx 40^\circ$. The appearance of the model corresponds to the timing of the CHARIS observations taken on 2020 Sept 2 ($\varphi=0.467$) and assumes a constant dust expansion velocity of 2450 km s$^{-1}$. The orbital configuration of the WR and O stars and the direction of motion of the O star (in the frame of reference of the WR star) is also shown in the models, where the red region indicates the orbital range of active dust formation.
    The adopted true anomaly range of dust formation is $-135^\circ$ to $135^\circ$ \citep{Han2022}. The persistent circumstellar dust features are identified in the observed view of the geometric wind-wind interface model, where north is up and east is to the left.}
    \label{fig:Surface}
\end{figure*}

With the exception of the Western Arm, all three observations capture the same set of persistent dust features given their similar morphology and positions relative to the central source.
In the COMICS image, which was taken at $\varphi=0.444$ in WR~140's orbital phase, C0 and C1 were located $\sim2.5\times$ and $\sim3.0\times$ the 11.7 $\mu$m FWHM from the central source which corresponds to $\sim$0\farcs8 and $\sim$1\farcs0, respectively. The extent of the bar in the COMICS image is $\sim1$\farcs0. In both CHARIS and NIRC2 observations, which were taken at $\varphi=0.467$ and $\varphi=0.462$ in WR~140's orbital phase, respectively, C0 and C1 were located 0\farcs9 and 1\farcs1 from the central source, and the extent of the bar was 1\farcs2. 
The slightly increasing size and displacement of the dust features measured over the $\sim2$ months spanning the COMICS, CHARIS, and NIRC2 observations are due to the high proper motion of the expanding dust (240--320 mas yr$^{-1}$; \citealt{Williams2009})

\subsection{Comparison to Predicted Dust Morphology}

A geometric model of dust formed in the wind collision interface between the WR~140 binary \rev{components} around periastron passage can be derived from its well-defined orbital properties \citep{Monnier2011,Thomas2021} and the previously characterized half-opening angle, $\theta\approx 40^\circ$, of the colliding-wind shock from the central binary \citep{Williams2009,Fahed2011}. 
Similar geometric models have successfully reproduced the morphology of circumstellar dust formed in the colliding-wind WC binaries WR~104 \citep{Soulain2018}, WR~112 \citep{Lau2020b}\rev{,} and Apep \citep{Callingham2019,Han2020}. Figure~\ref{fig:Surface} shows the \rev{orbital-plane and on-sky projections} of the dust-forming wind-wind interface model of WR~140's dust shell assuming a constant dust expansion velocity of $2450$ km s$^{-1}$, which is consistent with the observationally derived expansion velocity from \citet{Williams2009}. 
The orbital parameters of WR~140 in the geometric models are adopted from the \citet{Monnier2011} results with the \citet{Fahed2011} prior: $e = 0.9$, $\omega = 46.8^\circ$, $\Omega = 353.6^\circ$, $i = 119.6^\circ$.
The orbital configuration of the central binary at an orbital phase of $\varphi=0.47$ and the orbital range of dust formation is also shown in Fig.~\ref{fig:Surface}. The adopted true anomaly range where dust formation occurs is $-135^\circ$ to $135^\circ$\rev{; however, note that relatively little dust is observed around anomaly ranges consistent with periastron passage due to variable dust-formation rates\citep{Han2022}.}

Figure~\ref{fig:Surface} (\textit{Right}) indicates the position of the C0 and C1 dust features as well as the bar and the Western and Eastern Arms. These features likely correspond to regions of the dust shell with larger integrated dust column densities along the line of sight. The morphology of the Eastern Arm and the Western Arm, the latter of which \rev{captured only} in the NIRC2 image, appear consistent with the east and west edges of the near side of the geometric model. Based on the geometric model, the southern edge of the near side of the dust shell (i.e.,~the southern extent of the Arms) is coincident with C0 along the bar.
Given the lower angular resolution of the 11.7 $\mu$m COMICS observations relative to \rev{those} of NIRC2 and CHARIS, the blended dust emission of those two regions in the dust shell may contribute to C0 appearing brighter than C1 in the COMICS image. 

A notable \rev{difference} between the geometric wind-wind interface model of dust formation and the observations of WR~140 is the absence of bright dust emission along the northern edge of the dust shell, where the integrated column density along the line of sight toward the observer should be comparable to that of southern region associated with the bar. This is likely due to WR~140's highly variable dust-formation \rev{rates} throughout periastron passage \citep{Han2022,Eatson2022b}. 
Further investigations comparing colliding-wind simulations (e.g.,~\citealt{Eatson2022a,Eatson2022b}) with observations are therefore important for testing our understanding of dust formation in colliding-wind binaries; however, this is beyond the scope of this paper.

\begin{figure*}[t!]
    \includegraphics[width=.98\linewidth]{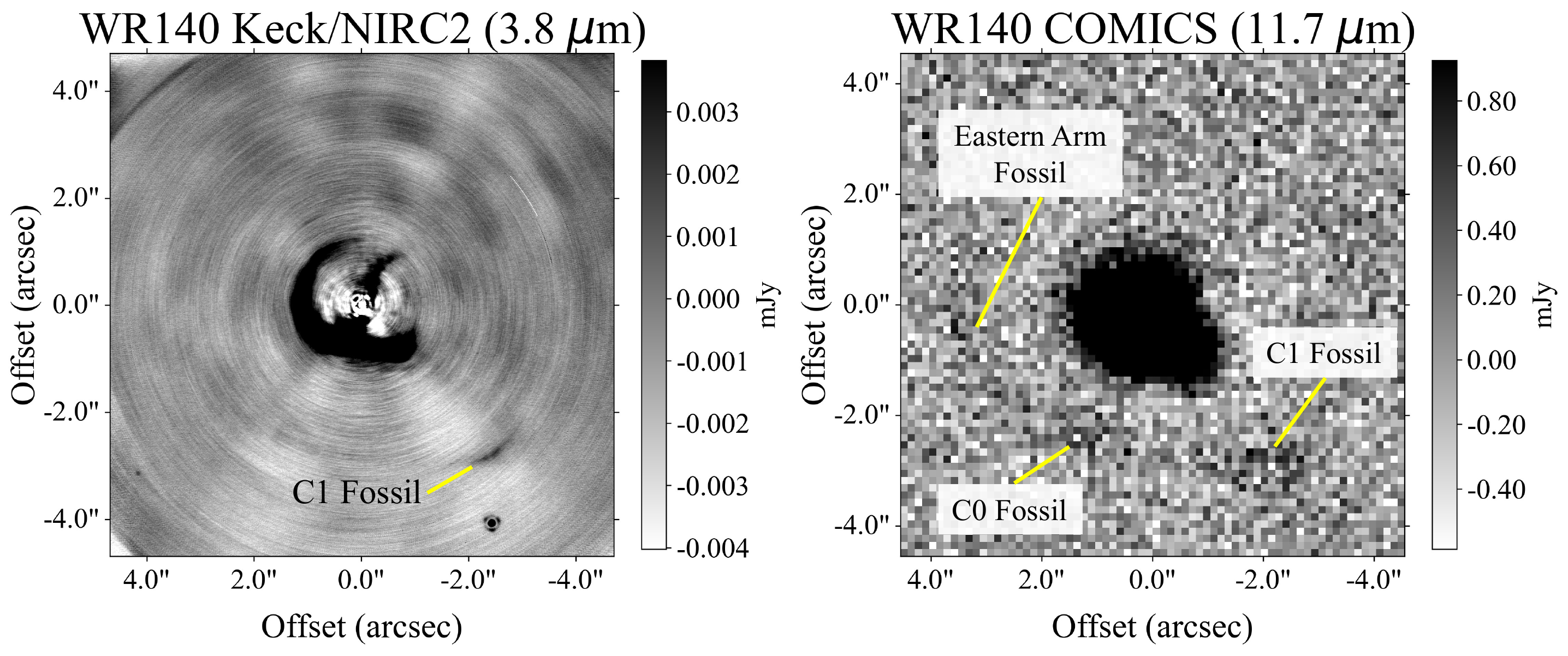}
    \caption{Keck/NIRC2 and Subaru/COMICS images of WR~140 displayed with a flux density range to highlight the detection of the C1 `Fossil' that formed during the 2009 periastron passage. The C0 Fossil and Eastern Arm Fossil are also detected in the COMICS image.}
    \label{fig:Full}
\end{figure*}

\subsection{WR~140 `Fossil' Dust Morphology from \rev{the} January 2009 Periastron Passage}
\label{sec:fossil}

The 3.8 $\mu$m NIRC2 and 11.7 $\mu$m COMICS images of WR~140 detect emission from the C1 feature that formed during the January 2009 periastron passage, one orbital period (7.93 yr) before the December 2016 periastron passage. Figure~\ref{fig:Full} shows the NIRC2 and COMICS images and the location of this C1 `Fossil' that is located along the same position angle as C1. The COMICS image also shows a faint detection of the C0 and Eastern Arm Fossils to the east of the C1 Fossil. The COMICS image resembles previous Gemini South imaging observations at 12.5 $\mu$m by \citet{Marchenko2007} and \citet{,Williams2009} \rev{as well as JWST imaging} where \rev{repeated} dust features from multiple \rev{past} dust-formation episodes \rev{were} detected. The NIRC2 image presents the first 3.8 $\mu$m detection of a previous dust-formation episode. 

In order to verify that the C1 Fossil in the NIRC2 image is indeed from the January 2009 dust-formation episode, the distance between C1 and the C1 Fossil can be compared to the expected separation distance based on previously measured proper motions of C1. The proper motion of C1 is 326 mas yr$^{-1}$ \citep{Williams2009}; therefore, over a 7.93 yr orbital period the separation distance between C1 and the C1 Fossil should be 2.6\arcsec, which is almost exactly the measured separation distance in the NIRC2 image.


\begin{deluxetable}{ccccc}
\tablecaption{WR~140 Dust Feature Observed Flux Densities}
\tablewidth{0pt}
\tablehead{$\lambda$ ($\mu$m) & $F_\mathrm{bar}$ & $F_\mathrm{C1}$ & $F_\mathrm{C1,Fossil}$ & Ext.~Cor}\
\startdata
1.241 & 3.62 & 2.26 & - & 2.26 \\
1.284 & 7.54 & 3.5 & - & 2.17 \\
1.329 & 8.04 & 4.21 & - & 2.09 \\
1.375\tablenotemark{*} & 9.09 & 5.1 & - & 2.02 \\
1.422\tablenotemark{*} & 11.83 & 5.43 & - & 1.95 \\
1.471 & 13.79 & 6.46 & - & 1.89 \\
1.522 & 15.34 & 7.19 & - & 1.83 \\
1.575 & 15.75 & 7.79 & - & 1.77 \\
1.63 & 18.83 & 9.5 & - & 1.72 \\
1.686 & 19.26 & 10.76 & - & 1.68 \\
1.744 & 22.93 & 12.39 & - & 1.64 \\
1.805 & 26.78 & 15.06 & - & 1.6 \\
1.867\tablenotemark{*} & 30.1 & 16.75 & - & 1.56 \\
1.932\tablenotemark{*} & 38.58 & 19.78 & - & 1.53 \\
1.999 & 42.78 & 21.47 & - & 1.5 \\
2.068 & 48.43 & 25.2 & - & 1.47 \\
2.139 & 59.82 & 29.02 & - & 1.44 \\
2.213 & 65.86 & 31.74 & - & 1.41 \\
2.29 & 69.44 & 32.67 & - & 1.39 \\
2.369 & 99.2 & 46.23 & - & 1.37 \\
3.8 & 385.26 & 169.47 & 10 & 1.17 \\
11.7 & 439.04 & 270.98 & 23 & 1.13 \\
\enddata
\tablecomments{Observed WR~140 flux densities (in mJy) of the bar, C1, and C1 Fossil. The extinction correction (Ext. Cor) column indicates the multiplicative derredening factor at each wavelength derived from the \citet{Gordon2021} interstellar extinction curve with $A_v=2.21$. The $1\sigma$ photometric uncertainties are $\sim10\%$ (See Sec.~\ref{sec:Obs}). \rev{The apertures used to measure the flux densities are described in Sec.~\ref{sec:Obs}.}}
\tablenotetext{*}{Telluric-dominated channel}
\label{tab:flux}
\end{deluxetable}

\subsection{The IR SED of WR~140 Dust and the Dust Temperature Disparities}
\label{sec:Tdisc}

The resolved mid-IR emission from circumstellar dust around WR~140 has been previously characterized and attributed to thermal emission heated radiatively by the central binary \citep{Williams2009}. \citet{Monnier2002} had first resolved near-IR emission around WR~140 \rev{with observations taken} shortly after periastron passage ($\varphi\approx0.04\text{--}0.06$) and suggested it originated from thermal emission from newly-formed dust. Based on $JHK$ light curves of WR~140 and the decreasing dust temperatures as dust drifts further from the central heating source, the near-IR emission from WR~140 was thought to fade away by around $\varphi\sim0.4-0.6$ \citep{Williams2009}. However, the 0.04 mag uncertainties in the $JHK$ magnitudes from the \citep{Williams2009} light curve of WR~140 indicate that faint near-IR emission could still persist at late times. Indeed, the CHARIS observations reveal the presence of near-IR circumstellar emission around WR~140 at an orbital phase greater than $\varphi\gtrsim0.4$, which highlights the importance in understanding the origin of the near-IR emission. The two most likely sources of circumstellar near-IR emission at $\varphi\sim0.45$, which is $\sim3.6$ yr after the 2016 Dec dust-forming episode, are thermal dust emission and scattered light from the central binary.

In order to investigate the properties of the spatially resolved near-IR emission, the spectral energy distribution (SED) of the bar and C1 were extracted from the CHARIS observations. 
The first two channels of the CHARIS data cube, which correspond to 1.16 and 1.20 $\mu$m, were heavily affected by data reduction artifacts and thus omitted from the SED analysis. The channels corresponding to 1.37, 1.42, 1.87, and 1.93 $\mu$m are likely telluric-dominated and were also omitted from the SED analysis.
The extracted near-IR flux densities \rev{were} corrected for interstellar extinction using the average observed extinction curve of the Milky Way ``G21\_MWAvg" from the ``dust\_extinction" Python package \citep{Gordon2021} \footnote{\url{https://github.com/karllark/dust_extinction}}. A value of $A_v=2.21$\footnote{The $v$ filter ($\lambda_\mathrm{C}=5160\,\AA$) is on the narrow-band system designed by \citet{Smith1968} where $A_v=1.1 A_\mathrm{V}$} was adopted for the interstellar extinction toward WR~140 \citep{Rate2020}.

\begin{figure*}[t!]
    \includegraphics[width=.49\linewidth]{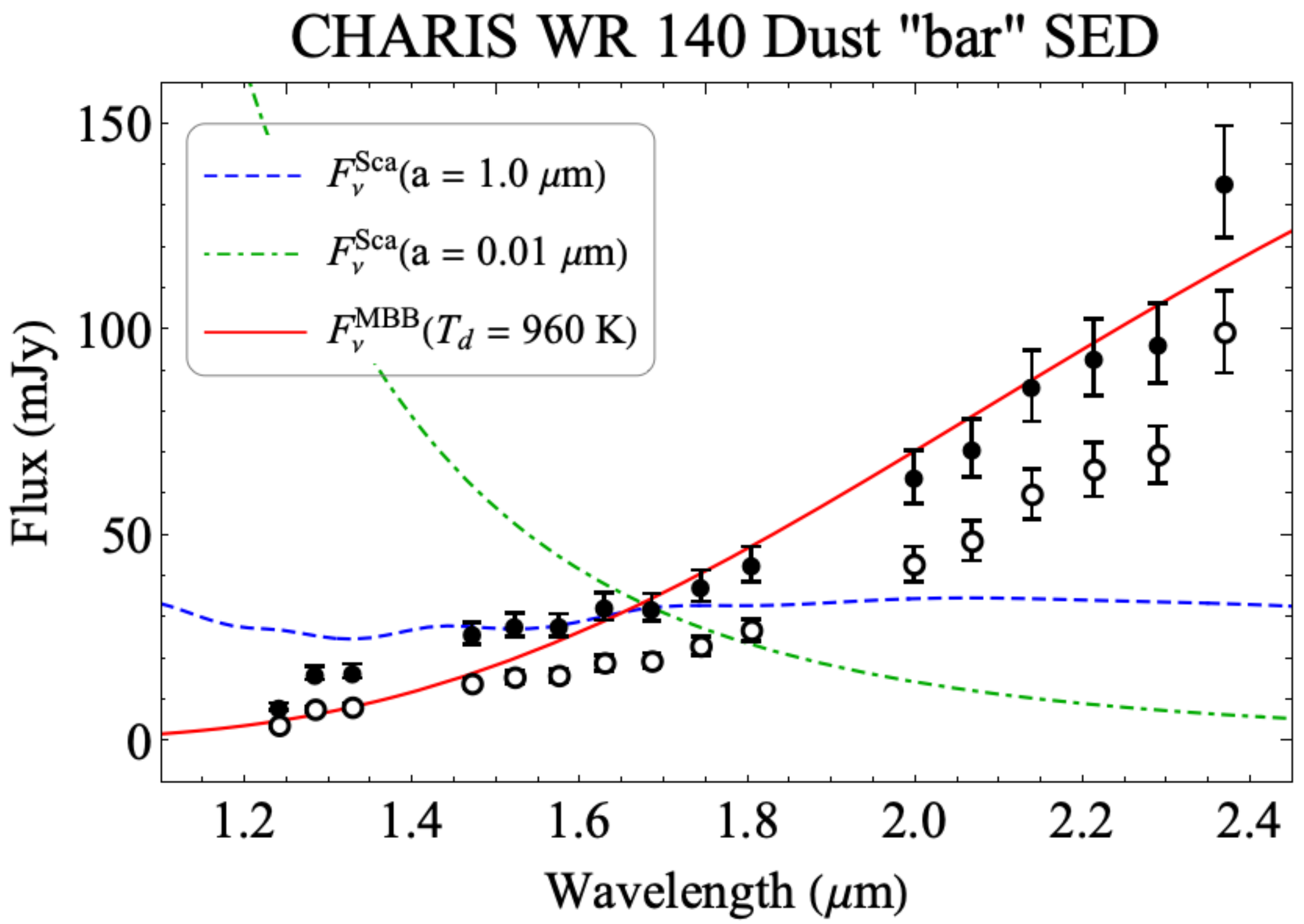}
    \includegraphics[width=.482\linewidth]{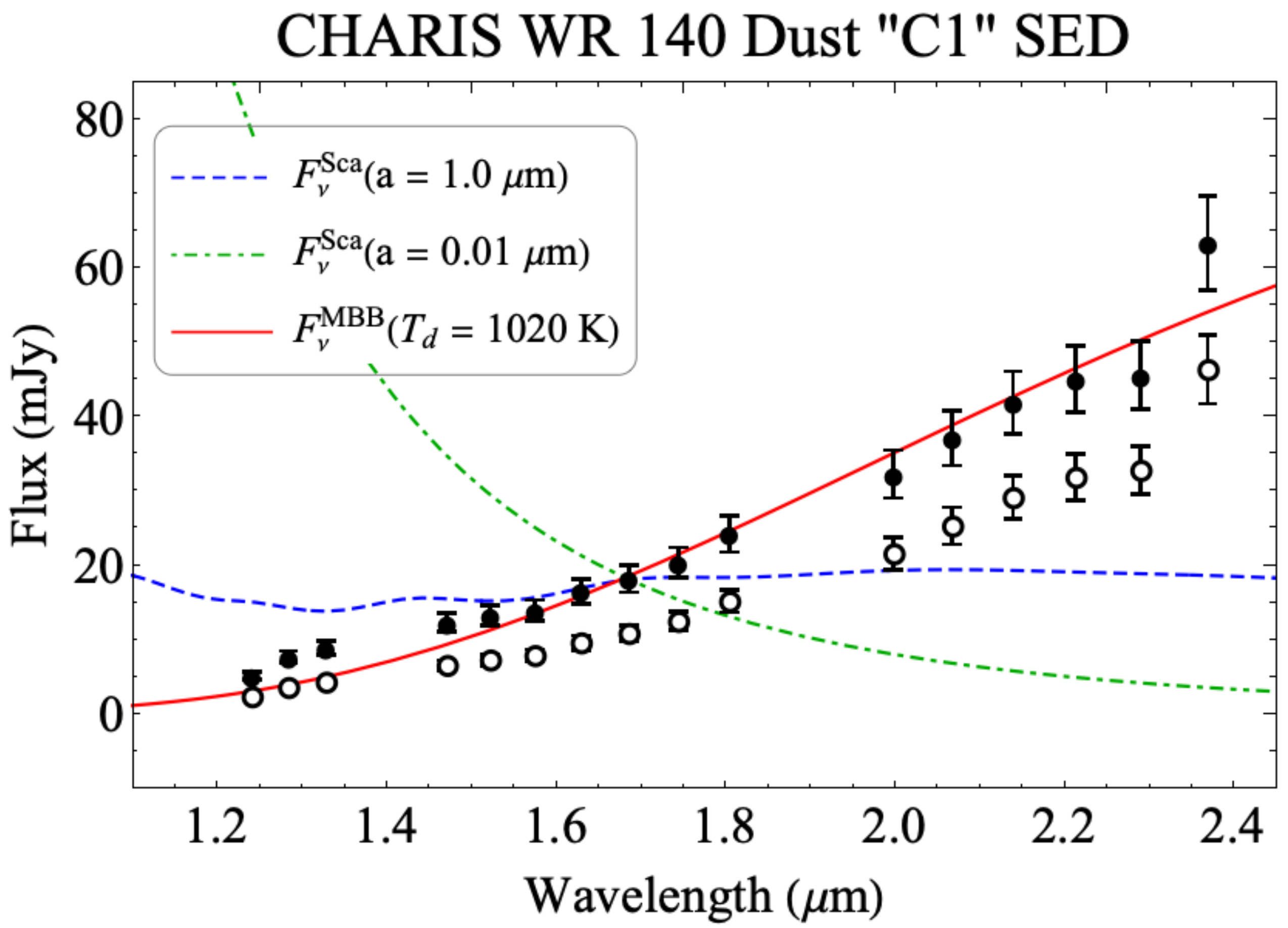}
    \caption{Observed (open circles) and \rev{interstellar-}extinction corrected (filled circles) 1.2--2.4 $\mu$m CHARIS \rev{SEDs} of WR~140 extracted from the (\textit{Left}) bar and the (\textit{Right}) C1 dust features. The \rev{SEDs} are overlaid with the scattered light models, \rev{which were normalized to the $H$-band flux density of the SEDs}, and \rev{the} modified black body emission model fit. The modified black body thermal dust emission model (Eq.~\ref{eq:MBB}) presents the best fit to the shape of the CHARIS \rev{SEDs} of both the bar and C1. Telluric dominated channels (See Tab.~\ref{tab:flux}) are omitted from the plots. }
    \label{fig:CHARIS}
\end{figure*}

Figure~\ref{fig:CHARIS} shows the near-IR SEDs of the bar and C1 from the CHARIS data cube before and after extinction correction. The shape of the near-IR SEDs of both the bar and C1 clearly demonstrates that the dust emission increases with increasing wavelength. 
\rev{Note that because the bar and C1 are spatially resolved from the central stars, it is unlikely that there is any contamination by emission from stellar winds in the SEDs.}
In order to understand the source of the near-IR flux from the bar, the two most likely scenarios were investigated: scattered light from the central binary or thermal dust emission. For the scattered light model, $F_\nu^\mathrm{Sca}(a) \propto Q_\mathrm{Sca}(a)\,F^*_{\nu}$, the scattering efficiencies Q$_\mathrm{Sca}(a)$ for aromatic-rich H-poor amorphous carbon (a-C) models with radii $a=0.01$ $\mu$m and $a=1.0$ $\mu$m were adopted \citep{Jones2013}. This \rev{a-C} dust model is also used \rev{for} SED modeling in Sec.~\ref{sec:2016SED}~\&~\ref{sec:2009SED}.

The near-IR continuum emission from WR~140 is assumed to be dominated by the free-free emission from the ionized stellar winds of the WC star, which can be approximated as a power-law $F_\nu^\mathrm{ff} \propto \nu^{0.96}$ \citep{Morris1993}. Both 0.01 and 1.0 $\mu$m grain size scattered light models, which were normalized to the $H$-band flux density of the SEDs, fail to reproduce the positive near-IR slope of SEDs and indicates that there is little or no contribution from scattered light (Fig.~\ref{fig:CHARIS}).

A modified black-body with an emissivity power-law index of $\beta=1$ was assumed for the thermal dust emission model,
\begin{equation}
F_\nu^\mathrm{MBB}(T_d) \propto B_\nu(T_d)\, \nu^\beta,
\label{eq:MBB}
\end{equation}
\noindent
where $T_d$ is the temperature of the emitting dust. The dust temperature and a multiplicative scaling factor for $F_\nu^\mathrm{MBB}(T_d)$ were fit to the CHARIS SEDs of the bar and C1. The best-fit thermal dust emission models with temperatures of $T_d\approx960$ K and $T_d\approx1020$ K for the bar and C1, respectively, demonstrate a close agreement with the shape of the near-IR SEDs. Even at late times ($\varphi\approx0.45$), the near-IR circumstellar emission around WR~140 is therefore most likely dominated by hot thermal dust emission.

Based on previous resolved mid-IR imaging studies of WR~140, the circumstellar 3.8 and 11.7 $\mu$m emission observed by NIRC2 and COMICS likely arises from thermal dust emission \citep{Williams2009}.
The observed near-IR, 3.8 $\mu$m, and 11.7 $\mu$m flux densities extracted from the bar and C1 are provided in Table~\ref{tab:flux} along with the extinction correction factors derived from the \citet{Gordon2021} interstellar extinction curve with $A_v=2.21$. Table~\ref{tab:flux} also provides the 3.8 and 11.7 $\mu$m flux densities measured from the C1 Fossil.

Adopting the same modified black-body thermal dust emission model fit to the near-IR emission (Eq.~\ref{eq:MBB}), the mid-IR 3.8 and 11.7 $\mu$m flux densities of the bar and C1 indicate temperatures of $T_d\approx560$ K and $T_d\approx530$ K, respectively. These temperatures are notably consistent with the circumstellar dust temperature derived from mid-IR observations by \citet{Williams2009} at a similar phase, where $T_d\approx560$ K.
The dust temperature of the C1 Fossil can also be estimated from its measured mid-IR 3.8 and 11.7 $\mu$m flux densities assuming they originate from the same dust component. Eq.~\ref{eq:MBB} provides a dust temperature of $T_d\approx490$ K for the C1 Fossil.

There are two notable disparities in the dust temperature calculations:

\begin{itemize}
  \item The near-IR thermal dust emission from the bar and C1 ($T_d\sim1000$ K) exhibits much hotter temperatures than those derived by the mid-IR 3.8 and 11.7 $\mu$m observations ($T_d\sim550$ K). 
  \item The derived temperature of the C1 Fossil ($T_d\approx490$ K) \rev{from the 3.8 and 11.7 $\mu$m observations} is a factor of $1.6$ hotter than predicted from the decreasing radiative equilibrium temperature as a function of distance from the heating source ($T_d\propto r^{-0.4}$).
\end{itemize}
Since the resolved circumstellar emission components at near- and mid-IR wavelengths are largely co-spatial, it is highly unlikely that the first disparity is due to spatial variations in the exposure to the central binary's radiation field that dominates the dust heating. 
This disparity between the near- and mid-IR dust temperatures of the bar and C1 instead suggests that the dust components probed by near- and mid-IR wavelengths possess different physical properties such as grain size.

\rev{The second disparity arises since dust continuum emission in radiative thermal equilibrium with a central heating source should decrease with distance (e.g.,~\citealt{Williams2009,Lau2020a}). The C1 Fossil is located over a factor 3.8 further than C1 from the central binary, which implies the dust temperature of the C1 Fossil should be $\sim310$ K given the 530 K temperature measured for C1 and adopting $T_d\propto r^{-0.4}$ for dust in equilibrium heating. This is significantly lower than the estimated 490 K dust temperature from the 3.8 and 11.7 $\mu$m flux measurements from Eq.~\ref{eq:MBB}.} The second disparity \rev{can also} be resolved if there are distinct dust components probed by the 3.8 and 11.7 $\mu$m observations at the C1 Fossil. This scenario at the C1 Fossil would invalidate the dust temperature estimate from Eq.~\ref{eq:MBB}, which assumed the 3.8 and 11.7 $\mu$m emission originates from the same dust component. These disparities are investigated in a detailed dust SED analysis in the following sections.

\begin{deluxetable}{lc}
\tablecaption{\dustem~SED Parameters}
\tablewidth{0pt}
\tablehead{Property & Value}\
\startdata
Grain Type& Aromatic-rich H-poor a-C\\
Bulk Density & 1.6 g cm$^{-3}$\\
Size Distribution & log-normal\\
$a_{min}-a_{max}$ & $9\times10^{-4}\text{--}1.0$ $\mu$m\\
$r_{dust}$ & 1650 au\\
$r_{C1, Fossil}$ & 6200 au\\
$L_{*,WR}$ & $9.3\times10^5$ L$_\odot$\\
$T_{2/3,WR}$ & 57,000 K\\
$R_{2/3,WR}$ & 9.79 R$_\odot$\\
$L_{*,O}$ & $5.0\times10^5$ L$_\odot$\\
$T_{2/3,O}$ & 36,000 K\\
$R_{2/3,O}$ & 18.2 R$_\odot$\\
$d_\mathrm{WR140}$\tablenotemark{a} & 1.64 kpc\\
\hline
\multicolumn{2}{c}{SED Grid Search Parameters}\\
\hline
$a_0$ (Nanodust) & 9--13 \AA~($\Delta a = 0.5$ \AA)\\
$\sigma$ (Nanodust) & 0.1\\
$a_0$ (Larger Grains) & 50--1100 \AA~($\Delta a = 100$ \AA)\\
$\sigma$ (Larger Grains)& 1.0\\
\enddata
\tablecomments{The aromatic-rich H-poor amorphous carbon (a-C) models are defined in \citet{Jones2013}. The parameters $a_\mathrm{min}$, $a_\mathrm{max}$, $r_\mathrm{dust}$, and  $r_\mathrm{C1,Fossil}$ correspond to the minimum and maximum grain sizes in the log-normal grain size distribution, the distance between the circumstellar dust and the central heating source, and the distance between the C1 Fossil and the central heating source, respectively. The heating source parameters $L_*$, $T_{2/3}$, and $R_{2/3}$ correspond to the luminosity, temperature (at $\tau=2/3$), and radius (at $\tau=2/3$) of the stellar heating sources, respectively. The SED grid search parameters show the range of the centroid grain sizes $a_0$ and step intervals $\Delta a$ for the nanodust and larger grain dust components in the \dustem~model fitting}
\tablenotetext{a}{\textit{Gaia}-derived distance to WR~140 \citep{Rate2020}}
\label{tab:SED_Param}
\end{deluxetable}

\begin{figure*}[t!]
    \includegraphics[width=.49\linewidth]{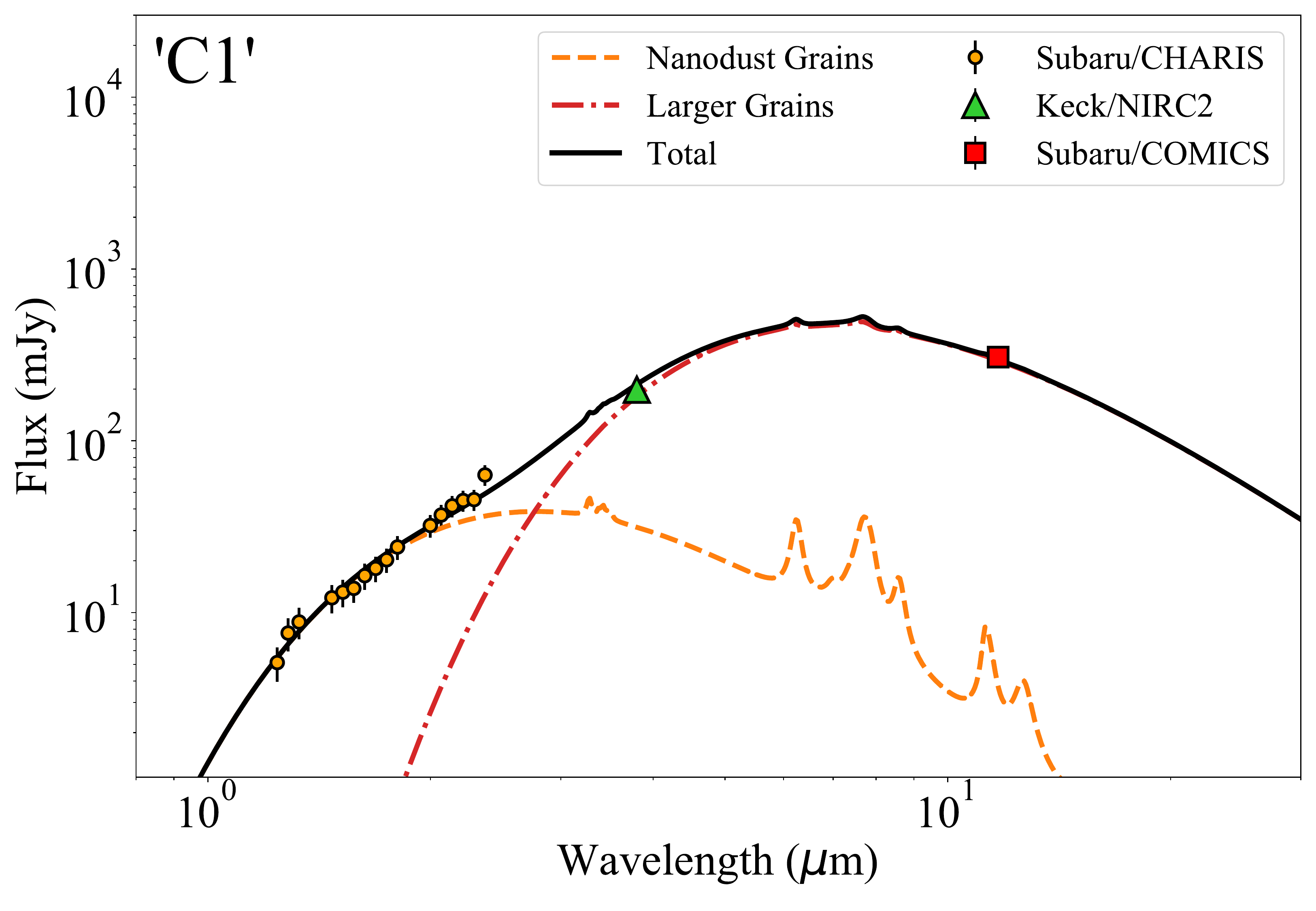}
    \includegraphics[width=.49\linewidth]{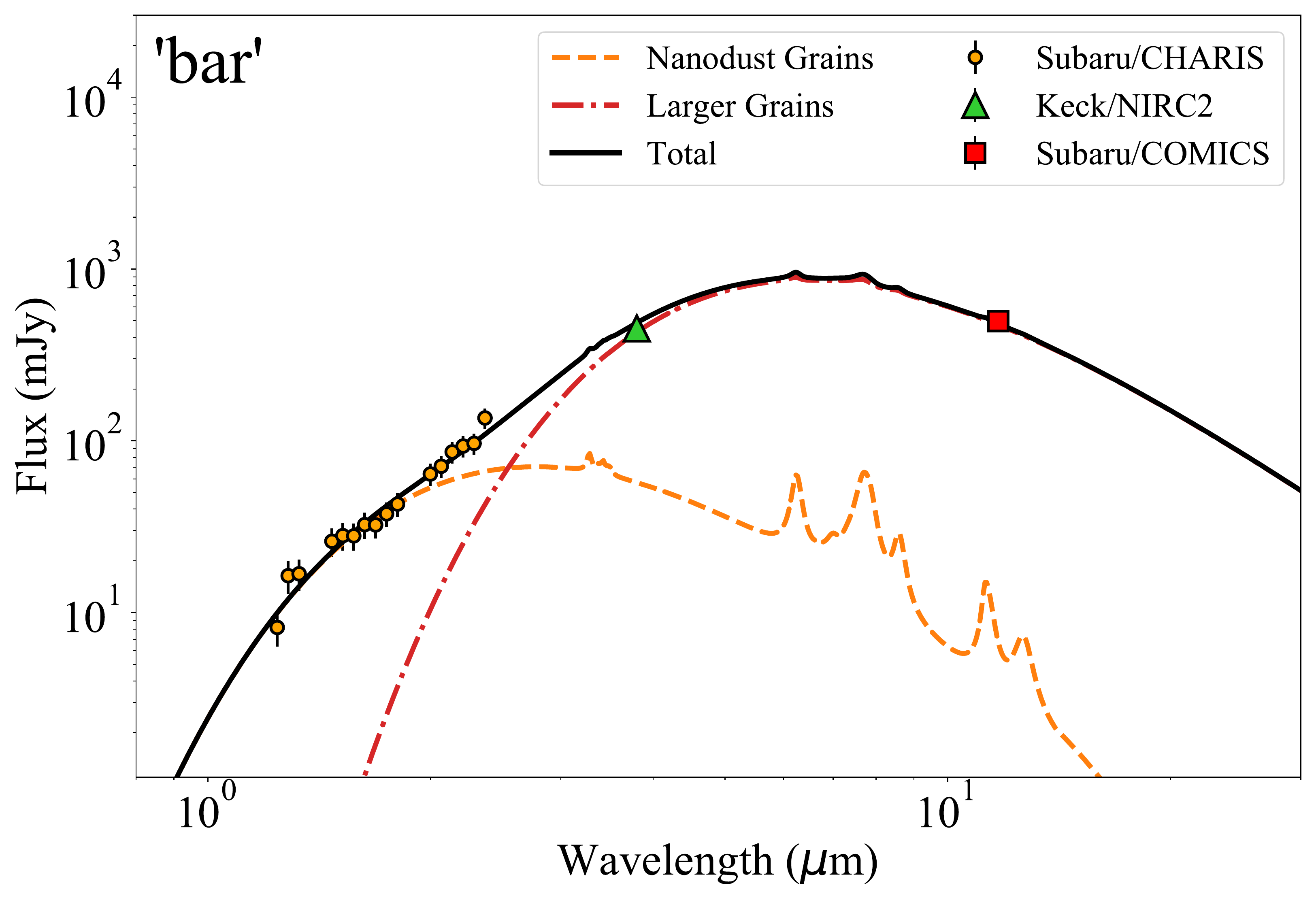}
    \caption{Best-fit nanodust (orange dashed line) + larger grain (red dot-dashed line) \dustem~SED models fit to the extinction-corrected CHARIS, NIRC2, and COMICS flux densities of circumstellar dust features (\textit{Left}) C1 and (\textit{Right}) the bar. The solid black line shows the combined emission from the nanodust and larger grain components. Telluric dominated channels from the CHARIS observations (See Tab.~\ref{tab:flux}) are omitted from the plots and SED fitting.}
    \label{fig:Fit}
\end{figure*}

\subsection{Dust SED Fitting of the 2016 Dust-Formation Episode: Revealing the Presence of Nanodust}
\label{sec:2016SED}

In order to investigate the properties of the near- and mid-IR dust components of the circumstellar dust around WR~140, dust emission models were fit to the 1.2--11.7 $\mu$m SEDs of the bar and C1 using \dustem. \dustem~is a numerical tool that computes the emission from dust heated by an input radiation field in the optically thin limit with no radiative transfer \citep{Compiegne2011}. Using the formalism of \citet{Desert1986}, \dustem~also calculates grain temperature distributions which is important for characterizing the emission from very small ($a\lesssim 10$ \AA) dust grains that are stochastically heated by single photon absorption (e.g.,~\citealt{Draine2001}). 
Unlike larger $a\gtrsim 100$ \AA~grains, these very small ``nanodust'' grains may not be in local thermal equilibrium (LTE) with the radiation field given their smaller absorption cross sections and lower heat capacities. Single photon absorption by nanodust grains that are not in LTE result in transient temperature spikes exceeding the temperatures of larger grains in LTE and can therefore produce an excess in shorter wavelength IR emission relative to the larger grains \citep{Sellgren1983, Desert1986,Guhathakurta1989}.
Notably, due to their lower heat capacities, smaller grains exhibit higher temperatures than larger grains when exposed to the same radiation field.
Emissions from such nanodust grains are likely to be important in explaining the hot temperatures of the near-IR component in WR~140. 

The input parameters for our \dustem~SED models were the radiation field of the central binary, the radial distance of the dust components from the central binary, the dust mass, and the grain size distribution properties. A log-normal grain size distribution with $dn/d\mathrm{log}(a) \propto e^{-(\mathrm{log}(a/a_0)/\sigma)^2}$ was adopted for the \dustem~models, where the free parameters were the centroid grain size $a_0$ and the distribution width $\sigma$. Minimum and maximum grain sizes in the models were fixed at $9\times10^{-4}$ $\mu$m and $1.0$ $\mu$m, respectively, where the minimum grain size limit was defined by where the maximum dust temperatures exceeded a sublimation temperature of $\sim1800$ K when exposed to the WR~140 radiation field.

To approximate the dust-heating radiation field from the WC7 and O5 stars in WR~140, we used synthetic spectra from model atmospheres computed with the PoWR code \citep{Grafener2002,Sander2012,Hainich2019}. The models were calculated to reproduce the spectral appearance of both stars (including the light ratio) using a detailed binary analysis similar to those described in \citet{Shenar2019}. A consistent treatment of the hydrostatic part is included in both the O and the WR model \citep{Sander2015}. A more dedicated paper about the spectral modeling of WR~140 is currently in the works (Shenar et al., in prep.). The derived effective temperatures at an optical depth of $\tau=2/3$ are $T_{2/3,\text{WR}}= 57,000$\,K and $T_{2/3,\text{O}}= 36,000$\,K, and the luminosities are $L_{*,\text{WR}}= 1\times10^6$\,L$_\odot$ and  $L_{*,\text{O}}=5\times10^5$\,L$_\odot$.

\begin{deluxetable*}{@{\extracolsep{8pt}}lcccc}
\tablecaption{WR~140 DustEM SED Properties of C1 and the bar}
\tablewidth{0pt}
\tablehead
{
\colhead{}&
  \multicolumn{2}{c}{C1}&
  \multicolumn{2}{c}{bar} \\
\cline{2-3} \cline{4-5} 
\colhead{Property}& \colhead{Larger Grains}& 
\colhead{Nanodust Grains}& \colhead{Larger Grains}& 
\colhead{Nanodust Grains}
}
\startdata
Grain Size Dist. Centroid ($a_0$) & $530^{+490}_{-210}$ \AA & $10.5^{+2.0}_{-0.5}$ \AA & $320^{+360}_{-200}$ \AA & $10.0^{+2.5}_{-1.0}$ \AA* \\
Grain Size Dist. Width ($\sigma$) & $1.0$  & $0.1$ & $1.0$  & $0.1$ \\
Dust Temp.~($T_{d}$) & 480 K  & 930 K & 520 K  & 930 K \\
IR Luminosity ($L_\mathrm{IR}$) & $21.5^{+1.2}_{-3.3}$ L$_\odot$ & $4.4^{+1.2}_{-0.3}$ L$_\odot$ & $42.6^{+3.3}_{-7.0}$ L$_\odot$ & $8.0^{+2.8}_{-2.3}$ L$_\odot$ \\
Mass ($M_d$) & $(2.4^{+1.6}_{-0.9})\times10^{-9}$ M$_\odot$ & $(1.1^{+0.3}_{-0.1})\times10^{-10}$ M$_\odot$ & $(3.2^{+2.0}_{-1.3})\times10^{-9}$ M$_\odot$ & $(2.1^{+0.7}_{-0.6})\times10^{-10}$ M$_\odot$ \\
$M_\mathrm{d,Tot}$  &\multicolumn{2}{c}{$(2.5^{+1.6}_{-0.9})\times10^{-9}$ M$_\odot$} & \multicolumn{2}{c}{$(3.4^{+2.0}_{-1.3})\times10^{-9}$ M$_\odot$}\\
$M_\mathrm{Nd}$/$M_\mathrm{d,Tot}$  &\multicolumn{2}{c}{$0.05^{+0.03}_{-0.02}$}& \multicolumn{2}{c}{$0.07^{+0.05}_{-0.03}$}
\vspace{4.0pt}
\enddata
\tablecomments{Best-fit \dustem~SED model parameters of WR~140's circumstellar C1 and bar dust features. The dust temperature $T_\mathrm{d}$ corresponds to the average dust temperature ``$T_\mathrm{moy}$'' output by \dustem~at the best-fit grain size centroid $a_0$. $L_\mathrm{IR}$ is the integrated IR luminosity of the larger grain or nanodust components. The individual dust masses ($M_\mathrm{d}$) of the larger and nanodust grains of the best-fit models are provided as well as the total ($M_\mathrm{d,Tot}$) dust masses of both components. $M_\mathrm{Nd}/M_\mathrm{d,Tot}$ corresponds to the dust mass ratio of the nanodust component to the total dust mass.
\\ *The lower uncertainty of $a_0$ for the nanodust grain \rev{population} of the bar hits the limit of the grid search.}
\label{tab:dustem}
\end{deluxetable*}

\rev{Based on the H-poor environment around WR stars and the recent claim of C-rich aromatic compounds in the circumstellar dust around WR~140 \citep{Lau2022},} dust grains are assumed to be composed of aromatic-rich H-poor amorphous carbon \rev{for SED modeling}. 
\rev{The properties of these grains} are consistent with the aromatic-rich H-poor carbonaceous grains defined in \citet{Jones2013}, which have a bulk density of $\rho_b=1.6$ g cm$^{-3}$. 
Notably, given its strong UV-radiation field, the central WR star is capable of rapidly photo-processing ($\tau_\mathrm{UV,pd}\ll 1$ yr; Eq.~7 from \citealt{Jones2014}) newly formed carbonaceous dust as large $a\lesssim 1000$ \AA~to a stable and fully evolved state consistent with the aromatic-rich H-poor amorphous carbon grains described by \citet{Jones2013}.
The distance to the circumstellar dust from the central binary system can be approximated by the spatially resolved imaging and the geometric  model (Fig.~\ref{fig:Collage}~\&~\ref{fig:Surface}). Adopting the \textit{Gaia}-derived distance to WR~140 of $d = 1.64$ kpc \citep{Rate2020}, the distance to the bar and C1 is $r_\mathrm{dust}\approx1650$ au and the distance to the C1 Fossil is $r_\mathrm{C1,Fossil}\sim 6200$ au \rev{at the time of the observations}.

\dustem~SED model fits to the bar and C1 were initially attempted using a single \rev{population} of $a\approx100$ \AA~dust grains, which was motivated by the $100$ \AA~grain size estimates from previous studies of circumstellar dust around WR~140 \citep{Williams2009,Lau2020a}. However, the single \rev{population} fits failed to reproduce the emission at both near- and mid-IR wavelengths. A second \rev{population} of \rev{$\sim10$ \AA}, nano-sized grains (i.e.,~``nanodust") was then introduced in the SED fitting. The best-fit \dustem~SED models were obtained from a reduced $\chi^2$ analysis, where the amplitude of the dust emission of the two components is modulated by their dust mass, and a grid search through two parameters: the grain size centroid of the nanodust and larger grain distributions.
Due to degeneracies with the grain size centroid,
the nanodust and larger grain size distributions' widths were fixed at values of 0.1 and 1.0, respectively.
The \dustem~SED grid search parameters and the other \dustem~model parameters are provided in Table~\ref{tab:SED_Param}.

\begin{figure}[t!]
    \includegraphics[width=.99\linewidth]{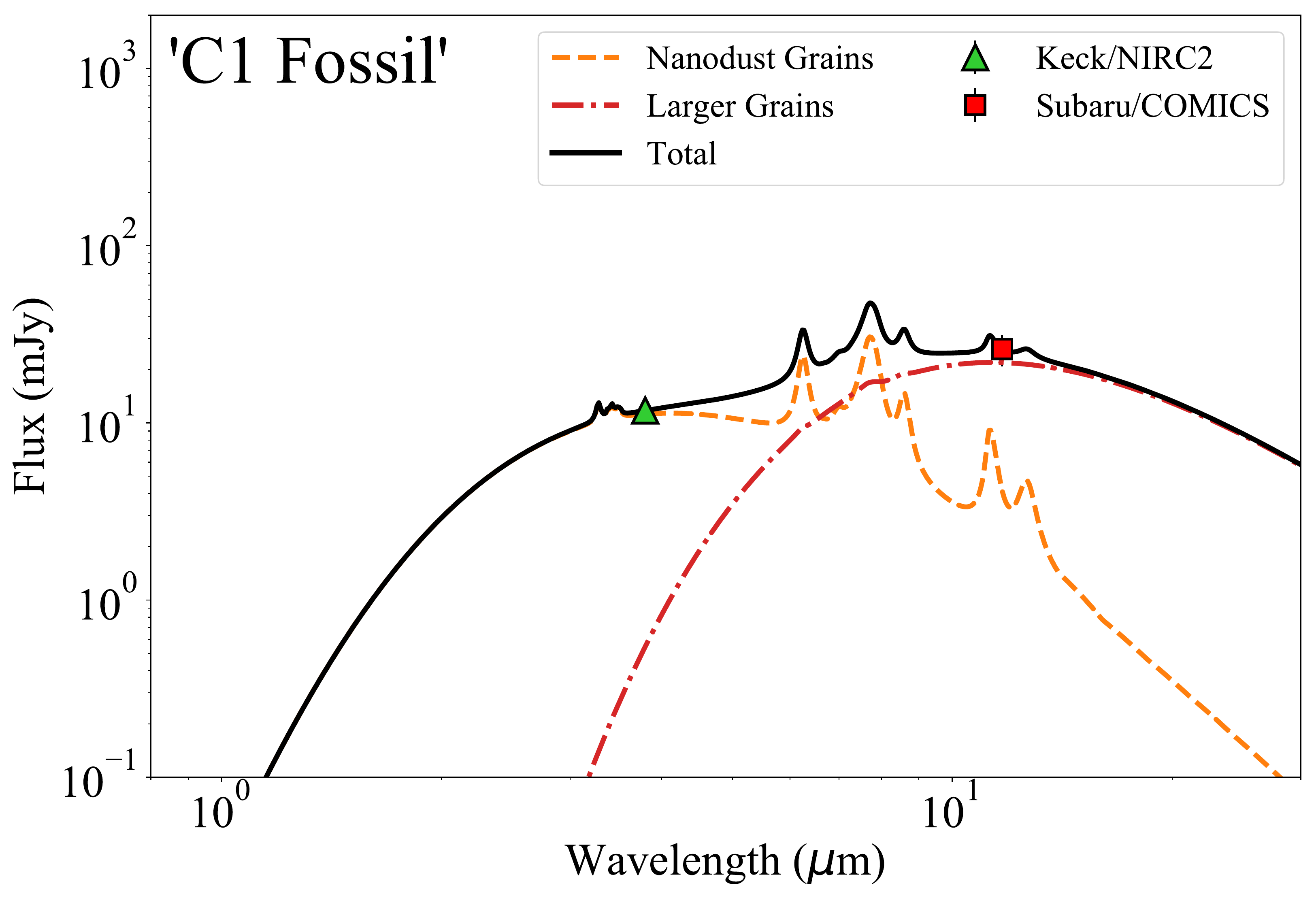}
    \caption{\rev{C1 Fossil \dustem~SED model of} nanodust (orange dashed line) + larger grain (red dot-dashed line) \rev{emission assuming the C1 Fossil dust is located a distance of $\sim6200$ au from the central binary with the grain size distribution centroids ($a_0$) of the nanodust and larger grains from the C1 SED model (Tab.~\ref{tab:dustem}). Similar to Fig.~\ref{fig:Fit}, the NIRC2 and COMICS flux densities of the C1 Fossil have been  corrected for interstellar extinction.} The solid black line shows the combined emission from the nanodust and larger grain components.}
    \label{fig:Fit_C1}
\end{figure}

The best-fit \dustem~SED models for the emission from the bar and C1 are shown in Figure~\ref{fig:Fit}. Both of the SED model fits demonstrate that the near-IR flux observed by CHARIS is dominated by thermal emission from $\sim10$ \AA~nanodust. \rev{Emission from the nanodust component between 3--14 $\mu$m also exhibits IR emission bands that are consistent with ``unidentified infrared (UIR)'' features found in various astrophysical environments \citep{Leger1984,Allamandola1985,Tielens2008}. Although the UIR features from dust around WR~140 were not covered by the ground-based observations presented in this work, they have been identified from mid-IR integral-field unit spectroscopy with JWST \citep{Lau2022}.}

The mid-IR flux observed by NIRC2 and COMICS is dominated by thermal emission from larger $\sim300\text{--}500$ \AA~grains \rev{in the best-fit SED models.} Although consistent within uncertainties, it is interesting to note that the grain sizes of this larger \rev{population} lie in between the $\sim100$ \AA~grains inferred from the analysis in \citet{Williams2009,Lau2020a} and the $\sim700$ \AA~grains determined by \citet{Marchenko2003} from optical photometric \rev{absorption} around periastron passage.  The dust SED properties of the best-fit models are provided in Table~\ref{tab:dustem}.

The dust temperature\footnote{$T_d$ corresponds the average dust temperature output by \dustem, ``$T_\mathrm{moy}$''.}, $T_d$, of the nanodust and larger grain size \rev{populations} are $930$ K and $\sim500$ K, respectively, and are consistent with the modified blackbody temperatures estimated in Sec.~\ref{sec:Tdisc}.
The total integrated IR luminosity of C1 and the bar are a very small fraction of the luminosity of the central binary ($L_{\mathrm{IR}}/L_*<0.01\%$). 

The total dust mass derived for the bar is $\sim3\times10^{-9}$ M$_\odot$. As expected, this value is slightly lower than the total WR~140 circumstellar dust mass ($6.4^{+3.8}_{-3.3}\times10^{-9}$ M$_\odot$) from similar SED modeling results from \citet{Lau2020a}, which was fitted to archival data taken at similar orbital phase ($\varphi\approx0.43$). 
In both C1 and the bar, the ratio of the nanodust mass to the total dust mass is $\sim5\%$. The general agreement between the dust temperatures and nanodust mass ratios of C1 and the bar suggests that C1 does not \rev{possess} significantly different dust properties from the other regions in the bar.

\subsection{Dust SED of the 2009 Dust-Formation Episode: Evidence of Continued Nanodust Production}
\label{sec:2009SED}

\rev{In order to estimate changes in mass from dust formed in the 2016 and 2009 dust-production episodes, the} \dustem~SED model \rev{fit to the C1 feature was applied} to the 3.8 and 11.7 $\mu$m detections of the C1 Fossil (Fig.~\ref{fig:Full}). Several free parameters \rev{were therefore} fixed due to the limited wavelength coverage of the C1 Fossil: the grain size distribution centroids ($a_0$) of the nanodust and larger grains was adopted from the derived C1 values (Tab.~\ref{tab:dustem}). The same heating source properties as the C1 and bar models were used. The separation distance between the central binary and the C1 Fossil was estimated from the NIRC2 observations to be $\sim6200$ au, which resulted in cooler dust temperatures for the larger and nanodust grains in the C1 Fossil. The only free parameters were therefore the dust masses of the nanodust and larger grain size \rev{populations, which were derived by scaling the nanodust and larger grain models to the 3.8 and 11.7 $\mu$m flux measurements.} 

The \rev{scaled} SED model of the C1 Fossil shown in Fig.~\ref{fig:Fit_C1} and the dust properties are provided in Table~\ref{tab:C1}. The dust SED fit suggests that the 3.8 $\mu$m emission is dominated by nanodust while the 11.7 $\mu$m emission is dominated by larger grains. Note that in the C1 and bar SED models (Fig.~\ref{fig:Fit}), the emission at both 3.8 and 11.7 $\mu$m emission was dominated by the larger grains. The different dust \rev{population} probed by the 3.8 $\mu$m emission in the C1 Fossil is \rev{likely} due to the cooler dust temperatures as well as an enhancement in the abundance of nanodust. The cooler dust temperatures of 300 K and 590 K exhibited by the larger grains and nanodust grains, respectively, in the C1 Fossil is expected due to $r^{-2}$ decrease in the radiative flux from the central binary heating source. \rev{A potentially} intriguing result is the increase in the C1 Fossil nanodust mass relative to the C1 dust mass by a factor of $\sim3$ and the decrease of the C1 larger grain dust mass by a factor $\sim0.4$. The ratio of the nanodust mass to the total dust mass in the C1 Fossil, $\sim40\%$, presents a substantial increase from the nanodust mass ratio in C1 of $\sim5\%$.

The total dust mass in the C1 Fossil, $\sim1.3\times10^{-9}$ M$_\odot$, is approximately half that of the total dust mass in C1, which \rev{suggests} that circumstellar dust around WR~140 is destroyed as it propagates away from the central binary. This is consistent with the interpretation and observations of the decreasing circumstellar dust mass by \citet{Williams2009} from their multi-epoch mid-IR imaging of WR~140. The $\sim4$ $\mu$m imaging observations presented by \citet{Williams2009}, however, did not capture the fossil dust emission, which was likely due to the limiting sensitivity of the observations. The NIRC2 observations of WR~140 therefore uniquely present evidence of an increasing abundance of nanodust. If the nanodust population is enhanced by the disruption of the larger grain population, then $\sim20\%$ of the total mass removed from the larger grain population between C1 and the C1 Fossil must have been transferred to new nanodust grains. 
\rev{It is important to note that these interpretations are derived from only two flux density measurements and should therefore be treated with caution until verified by observations at higher sensitivity and with broader wavelength coverage. However, UIR features that are typically attributed to very small carbonaceous grains and/or complex hydrocarbons were observed from the C1 Fossil with JWST \citep{Lau2022}, which supports the presence of nanodust grains derived from our SED models. The observation of such features in addition to the detection of the C1 feature from the past $\sim17$ dust-formation episodes at 7.7 $\mu$m \citep{Lau2022} suggest that nanodust grains are indeed present and survive in the extended circumstellar environment around WR~140. }

\begin{deluxetable}{lcc}
\tablecaption{WR~140 DustEM SED Properties of the C1 Fossil}
\tablewidth{0pt}
\tablehead{Property & Larger Grains  &  Nanodust Grains}\
\startdata
$a_0$ & $530$ \AA & $10.5$ \AA \\
$\sigma$ & $1.0$  & $0.1$ \\
$T_{d}$ & 300 K  & 590 K \\
$L_\mathrm{IR}$ & $1.0$ L$_\odot$ ($0.05\times$) & $0.6$ L$_\odot$ ($0.1\times$)  \\
$M_d$ & $\sim9.4\times10^{-10}$ M$_\odot$ ($0.4\times$) & $\sim3.6\times10^{-10}$ M$_\odot$ ($3\times$) \\
$M_\mathrm{d,Tot}$   & \multicolumn{2}{c}{$\sim1.3\times10^{-9}$ M$_\odot$ ($0.5\times$)}\\
$M_\mathrm{Nd}$/$M_\mathrm{d,Tot}$   & \multicolumn{2}{c}{$\sim0.4$ ($8\times$)}\\
\enddata
\tablecomments{\dustem~SED model parameters of the C1 Fossil dust feature. 
The grain size centroid for the larger grain and nanodust grain models ($a_0$) were adopted from the best-fit values of the C1 SED model. The dust temperatures, T$_d$, of the C1 Fossil are lower than that of the C1 feature (Tab.~\ref{tab:dustem}) due to the larger separation distance of the C1 Fossil from the central binary (Tab.~\ref{tab:SED_Param}). The factors in the parentheses following $L_\mathrm{IR}$, $M_\mathrm{d}$, $M_\mathrm{d,Tot}$, and $M_\mathrm{Nd}/M_\mathrm{d,Tot}$ indicate the multiplicative difference from the values derived for the C1 dust feature (e.g.,~the best-fit dust mass of the nanodust grains for the C1 Fossil is a factor of $\sim3$ greater than that of C1). }
\label{tab:C1}
\end{deluxetable}

\section{Discussion}
\label{sec:disc}
\subsection{Nanodust Formation around WR~140}
In this section, we consider the possible mechanisms responsible for the apparent enhancement of nanodust and destruction of larger $\sim300 - 500$ \AA~grains in the circumstellar environment of WR~140. This process must occur over the 7.93-yr timescale (i.e., the WR~140 orbital period) and the 1650 -- 6200 au distance scale that characterize how dust propagates from C1 to the location of the C1 Fossil. Based on the increasing nanodust abundance, we can rule out several dust destruction mechanisms that would produce the opposite result and destroy nanodust grains more efficiently than larger dust grains. 
Thermal sublimation, which is caused by heating grains \rev{to, and beyond,} their sublimation temperature, would preferentially \rev{tend to destroy small, stochastically-heated grains} since they are heated to higher temperatures than larger grains when exposed to an identical heating source, this is due to their lower 
heat capacities

Thermal and non-thermal sputtering of dust grains by collisions with gas in the circumstellar environment of WR~140 would destroy grains on timescales proportional to the grain size (e.g.,~\citealt{Tielens1994}), which implies that $\sim10$ \AA~nanodust would be destroyed via sputtering $\sim50$ times faster than $\sim500$ \AA~grains.
Both thermal sublimation and (non-)thermal sputtering can therefore be ruled out as dominant mechanisms for enhancing nanodust abundance.

We now consider the two most likely dust processing mechanisms that can produce smaller dust grains from larger grains: shattering via grain-grain collisions and rotational disruption from radiative torques \citep{Hirashita2010,Asano2013, Hoang2019}. 
The shattering of larger grains via grain-grain collisions is believed to be an important mechanism for significantly altering grain size distributions and enhancing the population of small grains (e.g.,~\citealt{Jones1994,Jones1996,Hirashita2010,Asano2013}).
Another mechanism that can increase the abundance of small grains relative to large grains is dust disruption from the centrifugal stress within fast-rotating grains spun up by radiative torques. \citet{Hoang2019} claim that this radiative torque disruption (RATD) process should be particularly relevant in environments with strong radiation fields from sources such as massive stars and supernovae. 
The viability of grain-grain collisions compared to disruption from radiative torques in the circumstellar environment of WR~140 can be investigated by comparing the relevant timescales of each mechanism. We specifically focus on the C1 dust concentration and its derived properties for estimating and comparing the timescales. 

\textbf{\textit{Grain-grain collisions.}} For grain shattering, the mean timescale for collisions between two grains of size $a$ can be estimated as follows: 

\begin{equation}
    \tau_\mathrm{gg}=\frac{1}{\pi a^2 n_\mathrm{gr} v_\mathrm{gr}},
    \label{eq:gg}
\end{equation}

\noindent
where $n_\mathrm{gr}$ is the number density of dust grains and $v_\mathrm{gr}$ is the relative velocity of the grains. Given the complexity of estimating the relative velocity of circumstellar dust formed via colliding-winds, we adopt an upper limit of $v_\mathrm{gr}$ as the drift velocity of dust grains through gas, $v_\mathrm{drift}$ (i.e.,~$v_\mathrm{gr}<v_\mathrm{drift}$).
\rev{The drift velocity can be estimated from the observationally derived radial expansion velocity of the circumstellar dust around WR~140, $v_\mathrm{dust}\sim2600$ km s$^{-1}$ \citep{Lau2022}, and the terminal velocity of the WC star, $v_\infty\approx2900$ km s$^{-1}$ \citep{Williams1989}. Therefore, $v_\mathrm{drift}=v_\infty-v_\mathrm{dust}\sim300$ km s$^{-1}$. }

\begin{equation}
\rev{v_\mathrm{drift}=v_\infty-v_\mathrm{dust}\sim300 \,\mathrm{km}\, \mathrm{s}^{-1}}
    \label{eq:drift}
\end{equation}

The number density of dust grain in the circumstellar dust shell is difficult to measure given its complex 3-D morphology (Fig.~\ref{fig:Surface}). However, we can estimate an upper limit on the dust grain number density for the partially spatially resolved C1 dust feature assuming that its derived dust mass (Tab.~\ref{tab:dustem}) is concentrated in a sphere with a radius equivalent to half the FWHM of the angular resolution in the Keck/NIRC2 imaging observations at the 1.64 kpc distance to WR~140, $\Delta r_\mathrm{C1} \approx 60$ au. An upper limit of the dust grain number density in the C1 feature can be expressed as

\begin{equation}
   n_\mathrm{gr,C1} \lesssim \frac{M_\mathrm{d,C1}}{m_\mathrm{gr} \frac{4}{3} \pi {\Delta r_\mathrm{C1}}^3} \approx
   \frac{M_\mathrm{d,C1}}{( \frac{4}{3} \pi a^3 \rho_\mathrm{bulk})\,
   \frac{4}{3} \pi {\Delta r_\mathrm{C1}}^3},
    \label{eq:ngr1}
\end{equation}
\noindent
where $M_\mathrm{d,C1}$ is the dust mass of the larger grains in C1, $m_\mathrm{gr}$ is the mass of an individual larger grain, and $\rho_\mathrm{bulk}$ is the bulk density of the dust grain. Based on the adopted and derived SED model dust properties of C1, we derive the following value as an upper limit

\begin{multline}
   n_\mathrm{gr,C1} \lesssim 1.8\times10^{-6} \left(\frac{M_\mathrm{d,C1}}{2.4\times10^{-9}\,\mathrm{M}_\odot}\right)
    \left(\frac{\Delta r_\mathrm{C1}}{60\,\mathrm{au}}\right)^{-3}\\
    \left(\frac{\rho_\mathrm{bulk}}{1.6\, \mathrm{g}\,{\mathrm{cm}^{-3}}}\right)^{-1}
    \left(\frac{a}{500 \,\mathrm{\AA}}\right)^{-3}
    \,\mathrm{cm}^{-3}.
    \label{eq:ngr2}
\end{multline}

\noindent
A lower limit on the grain-grain collision timescale of \rev{$\tau_\mathrm{gg,C1}\gtrsim7.5$ yr} can then be derived by combining Eqs.~\ref{eq:drift}~\&~\ref{eq:ngr2} with Eq.~\ref{eq:gg}:

\begin{multline}
    \tau_\mathrm{gg,C1}\gtrsim\rev{7.5}
    \left(\frac{n_\mathrm{gr,C1}}{1.8\times10^{-6} \,\mathrm{cm}^{-3}}\right)^{-1}\\
    \left(\frac{v_\mathrm{drift}}{\rev{300}\,\mathrm{km}\,\mathrm{s}^{-1}}\right)^{-1}
    \left(\frac{a}{500 \,\mathrm{\AA}}\right)^{-2}
    \,\mathrm{yr}.
    \label{eq:tgg}
\end{multline}

Given that grain-grain collision timescales are less than the 7.93-yr orbital period of WR~140, it is \rev{possible} that grain-grain collisions may be causing the redistribution of mass from larger to smaller dust grains.

\textbf{\textit{Radiative torque disruption (RATD).}} In strong radiation fields ($U\gg1$) with mean wavelength $\bar{\lambda}$, the RATD mechanism can disrupt dust grains with radii larger than a critical grain size $a_\mathrm{disr}$ \citep{Hoang2019}. Utilizing the derivation of $a_\mathrm{disr}$ from Eq.~26 in \citet{Hoang2019} for environments where $a_\mathrm{disr}\lesssim\bar{\lambda}/1.8$, we can estimate $a_\mathrm{disr}$ for circumstellar dust around WR~140 as follows:

\begin{multline}
    a_\mathrm{disr} \simeq 84
    \left[\left(\frac{\gamma}{1.0}\right)^{-1}
    \left(\frac{L_\mathrm{*,WR+O}}{1.43\times10^6\,\mathrm{L}_\odot}\right)^{-1/3} \right.\\
    \left.\left(\frac{r_\mathrm{dust}}{1650\,\mathrm{au}}\right)^{2/3} 
    \left(\frac{\bar{\lambda}_\mathrm{WR+O}}{0.11\,\mu\mathrm{m}}\right)^{1.7} 
    \left(\frac{S_\mathrm{max}}{10^9\,\mathrm{erg} \,\mathrm{cm}^{-3}}\right)^{1/2}\right]^{\frac{1}{2.7}} \,\mathrm{\AA},
    \label{eq:adisr}
\end{multline}

\noindent
where $\gamma$ corresponds to the anisotropy of the radiation field (i.e.,~$\gamma = 1$ for unidirectional radiation sources), $L_\mathrm{*,WR+O}$ is the combined luminosity of the WR and O stars in WR~140, $r_\mathrm{dust}$ is the distance between the circumstellar dust and the central binary, $\bar{\lambda}_\mathrm{WR+O}$ is the mean wavelength of the radiation field from the central WR+O binary, and $S_\mathrm{max}$ is the maximum tensile strength of the grain material. 

In Eq.~\ref{eq:adisr}, we have adopted $S_\mathrm{max}=10^9$ erg cm$^{-3}$, which is a typical value for compact grains and is also consistent with polycrystaline graphite \citep{Hoang2019}; however, we note that strong materials such as monocrystaline graphite and diamond can exhibit tensile strengths of $S_\mathrm{max}\gtrsim10^{11}$ erg cm$^{-3}$ \citep{Hoang2019}. The calculation of $a_\mathrm{disr}$ in Eq.~\ref{eq:adisr} demonstrates that the larger grain size \rev{population} in our SED models of the bar and C1 ($a\sim300\text{--}500$ \AA; Tab.~\ref{tab:dustem}) satisfies the conditions for which RATD should occur.

The RATD disruption time for grains of size $a$ can be defined as the time required to spin-up a grain to its rotational disruption velocity. \citet{Hoang2019} provide the expression for the RATD disruption time in their Eq.~27, which we can apply to the circumstellar environment of WR~140 for dust located at $r_\mathrm{dust}=1650$ au (e.g.,~the C1 feature) from the central binary:

\begin{multline}
    \tau_\mathrm{RATD} \simeq 1.6
    \left(\frac{\bar{\lambda}_\mathrm{WR+O}}{0.11\,\mu\mathrm{m}}\right)^{1.7}
    \left(\frac{L_\mathrm{*,WR+O}}{1.43\times10^6\,\mathrm{L}_\odot}\right)^{-1}\\
    \left(\frac{a}{500 \,\mathrm{\AA}}\right)^{-0.7}
    \left(\frac{r_\mathrm{dust}}{1650\,\mathrm{au}}\right)^{2} 
    \left(\frac{S_\mathrm{max}}{10^9\,\mathrm{erg} \,\mathrm{cm}^{-3}}\right)^{1/2}\,
    \mathrm{days}.
    \label{eq:tratd}
\end{multline}

A comparison of the grain-grain collision and RATD timescales (Eq.~\ref{eq:tgg}~\&~\ref{eq:tratd}) indicates that $\tau_\mathrm{RATD}\ll \tau_\mathrm{gg}$ for circumstellar dust around WR~140 at C1 located $r_\mathrm{dust}=1650$ au from the central binary. 
Dust disruption from radiative torques by the strong radiation field of the central binary may therefore be the dominant grain disruption mechanism; however, such short RATD timescales suggest the larger grain population should be entirely disrupted by the time C1 reaches the position of the C1 Fossil, which disagrees with the presence of larger grains in the C1 Fossil (Tab.~\ref{tab:C1}, Fig.~\ref{fig:Fit_C1}).
It is possible that the maximum tensile strength of the dust grains ($S_\mathrm{max}$) around WR~140 is underestimated by factors of $\sim10\text{--}100$, which would result in increased RATD timescales by factors of $\sim3\text{--}10$ (Eq.~\ref{eq:tratd}). In the RATD calculations, we have also disregarded the possible effects of grain rotational damping (e.g., by gas collisions), which may mitigate the grain disruption from radiative torques \citep{Hoang2019}.
We also note that there are significant uncertainties in the grain density estimates for the grain-grain collision timescale calculation (Eq.~\ref{eq:ngr1}).
Ultimately, given the strong and hard radiation field around WR~140, we slightly favor RATD over grain-grain collisions as the dominant mechanism that increases the abundance of nanodust grains relative to the larger grains in WR~140.

\begin{figure}[t!]
    \includegraphics[width=.98\linewidth]{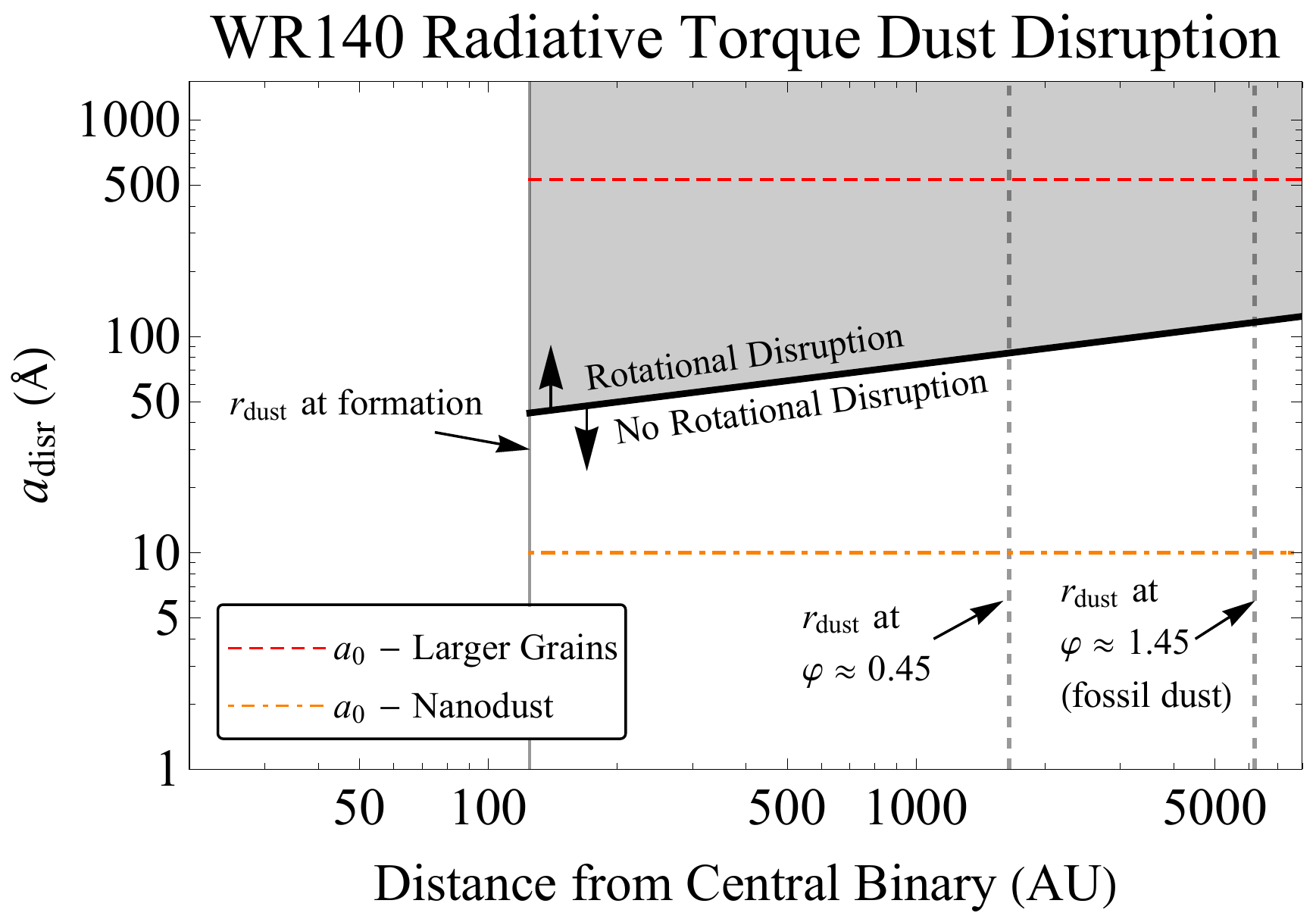}
    \caption{The critical size ($a_\mathrm{disr}$) for radiative torque disruption of dust around WR~140 as a function of distance from the central binary based on the analytical RATD expressions from \citet{Hoang2019}. Grains larger than $a_\mathrm{disr}$ will be disrupted by the RATD mechanism. The grain size centroids, $a_0$, of the nanodust and larger grain components fit to the circumstellar C1 and C1 Fossil dust features are overlaid as orange dot-dashed and red dashed lines, respectively. The separation distances from the central binary to the C1 and C1 Fossil features are also overlaid on the plot. }
    \label{fig:RATD}
\end{figure}

We can investigate the impact of RATD in the circumstellar environment of WR~140 starting from the point of dust formation at $r_\mathrm{dust}\sim125$ au \citep{Williams2009}. 
Figure~\ref{fig:RATD} shows the critical grain size $a_\mathrm{disr}$ as a function of $r_\mathrm{dust}$ from WR~140 based on Eq.~\ref{eq:adisr}. The $\sim10$ \AA~nanodust grains notably fall below $a_\mathrm{disr}$, which implies that RATD should not destroy newly formed nanodust created from disrupted larger grains. 
Interestingly, Fig.~\ref{fig:RATD} indicates that larger grains ($a\gtrsim100$ \AA) around WR~140 should continue to be disrupted by RATD at separation distances as large as the position of the C1 Fossil ($r_\mathrm{dust}\approx 6200$ au). The nanodust to total dust mass ratio should therefore increase beyond the $\sim40\%$ measured from our SED analysis of the C1 Fossil observations (Tab.~\ref{tab:C1}).

\subsection{Astrophysical Implications: An Early Source of Carbonaceous Nanodust?}
Owing to their rapid evolutionary formation timescales ($\sim$Myr) and high dust production rates ($\dot{M_d}\sim10^{-10}-10^{-6}$ M$_\odot$ yr$^{-1}$; \citealt{Lau2020a}), dust-forming WC binaries have been revisited in recent studies as early and important sources of interstellar dust \citep{Marchenko2017,Lau2020a,Lau2021a,Lau2022}.
The \rev{confirmation} of nanodust formation around a colliding-wind WC binary system like WR~140 therefore presents potentially significant implications on the origin of very small carbonaceous grains in the ISM of galaxies in the local and early Universe.
\rev{Here, we consider} whether or not all dust-forming WC binaries are capable of nanodust production like WR~140 \rev{based on the RATD mechanism.}

Newly formed dust in the circumstellar environment around all WC binaries should be exposed to a strong radiation field powered by the high luminosities ($\sim10^5-10^6$ L$_\odot$) and hot photospheric temperatures ($\gtrsim40,000$ K) of the central WC star as well as that of the \rev{OB}-star companion. Based on Eq.~\ref{eq:adisr}~\&~\ref{eq:tratd}, even the lowest luminosity WC9 stars ($L_*\sim10^5$ L$_\odot$) should be capable of disrupting grains via RATD that are larger than $a_\mathrm{disr}\sim100$ \AA~at a distance of $r_\mathrm{dust}=1650$ au on timescales as short as $\tau_\mathrm{RATD}\sim20$ days.
Interestingly, in their dust SED model fitting of Galactic dust-forming WC systems, \citet{Lau2020a} find that the \textit{episodic} dust-forming WC systems such as WR~140, WR~125, WR~137, and WR~19 are better fit by smaller grain size models ($a = 0.01\text{--}0.1$ $\mu$m) whereas the \textit{persistent} dust-forming systems such as WR~95, WR~98a, WR~104, WR~106, and WR~118 are better fit by larger grain size models ($a = 0.1\text{--}1.0$ $\mu$m). The dust production rates are higher in the persistent dust-forming systems compared to the episodic systems, which suggests that self shielding and larger optical depths could play a role in enhancing grain growth and mitigating the RATD mechanism in persistent dust makers. 

The \citet{Lau2020a} dust SED model fits of the persistent dust-forming WC systems that favor larger grains, however, do not preclude the presence of nanodust in these systems given the challenges of detecting nanodust emission. \rev{Indeed, previous mid-IR spectroscopic studies of dust-forming WC systems that included persistent dust-formers like WR~104 show that they exhibit UIR features and therefore indicate the presence of very small carbonaceous grains and/or complex hydrocarbons \citep{Chiar2002,Marchenko2017,Endo2022}.} Our observations of WR~140 notably demonstrate that high angular resolution multi-wavelength IR imaging observations are crucial for distinguishing nanodust from the longer wavelength IR emission by cooler and larger dust grains as well as the near-IR free-free emission from the ionized winds of the central WR star. 
We therefore speculate that all dust-forming WC binaries should exhibit nanodust production due to the RATD mechanism on larger grains. The nanodust production efficiency \rev{may} differ based on the dust-formation conditions around the WC binary. 
 
\rev{Given that the evolutionary timescales to form} WR stars ($\sim$Myr) is much shorter than that of AGB stars ($\gtrsim$ 100 Myr), which have been considered as leading sources of dust in the local Universe (e.g.,~\citealt{Gall2011}), \rev{we argue that WC binaries} present an early source of C-rich nanodust. 
\rev{We speculate that nanodust formed by} WC binaries \rev{are the C-rich aromatic carriers of the UIR} features detected from dusty WC binaries \citep{Endo2022, Lau2022}. \rev{Nanodust from WC binaries also} present a potential candidate as a carrier of the 2175 \AA~UV extinction bump, which is thought to arise from very small carbonaceous grains \citep{Stecher1965,Duley1998,Draine2009}.
\rev{Recent observations with JWST of galaxies from the first billion years of cosmic time notably present evidence of the 2175 \AA~absorption feature, which highlights the possible presence of carbonaceous dust in the early Universe \citep{Witstok2023}.}
Continued \rev{ground-based,} high contrast, and high angular resolution near/mid-IR observations \rev{in combination with sensitive spaced-based observations with JWST} will be essential for investigating the impact of nanodust formation in WC binaries.

\section{Conclusions}
In this work, we presented high contrast and high spatial resolution IR observations of the dust-forming colliding-wind binary WR~140 with Subaru/SCExAO + CHARIS, Subaru/COMICS, and Keck/NIRC2+PyWFS (Fig.~\ref{fig:Collage}~\&~\ref{fig:COMICS}). All observations were taken within $\sim2$ months and enabled a $1\text{--}12$ $\mu$m morphological analysis of the circumstellar dust formed in the most recent periastron passage and dust-production episode from WR~140 in December 2016. The persistent circumstellar dust features referred to as C0, C1, and the bar were present in all observations, but the dust feature referred to as the Eastern Arm was only detected at longer IR wavelengths (3.8 and 11.7 $\mu$m) by NIRC2 and COMICS. The imaging observations were compared against a geometric model of the dust-forming regions of WR~140's wind-wind interface at a consistent orbital phase ($\varphi=0.47$; Fig.~\ref{fig:Surface}). The observed morphology of the circumstellar dust emission appeared consistent with regions in the geometric model that should exhibit the largest integrated column density along our line of sight. However, other regions of enhanced column density predicted by the geometric model were not detected in the observations\rev{, which is likely due to variable dust production around periastron passage \citep{Han2022}}. The 3.8 $\mu$m NIRC2 and 11.7 $\mu$m COMICS images detected faint `Fossil' dust emission linked to the WR~140's dust-formation event during the January 2009 periastron passage.

An SED analysis of the IR circumstellar dust emission revealed that the spectral shape in the near-IR was consistent with thermal dust emission at a temperature of $\sim1000$ K as opposed to dust-scattered light from the central binary (Fig.~\ref{fig:CHARIS}). However, the $\sim1000$ K dust temperature derived from the near-IR emission was inconsistent with the $\sim550$ K dust temperatures determined from the 3.8 and 11.7 $\mu$m observations. Another circumstellar dust temperature disparity was identified: the $490$ K dust temperature of the C1 Fossil derived from the 3.8 and 11.7 $\mu$m images was a factor of 1.6 hotter than expected based on decreasing radiative equilibrium temperatures as a function of distance from the heating source. 

In order to investigate the dust properties and resolve the temperature disparities, we conducted a detailed dust SED analysis of the bar, C1, and the C1 Fossil with \dustem~(Fig.~\ref{fig:Fit}~\&~\ref{fig:Fit_C1}). The best-fit \dustem~SED models indicated the presence of a hot dust component composed of very small ($a\sim10$ \AA) nanodust grains. 
The presence of this hot dust component in addition to the previously known larger ($a\sim300\text{--}500$ \AA) and cooler dust component resolved the dust temperature disparities. Interestingly, the \dustem~SED model fit to the C1 Fossil \rev{indicated} a larger abundance of nanodust and a decreasing abundance of the larger grains in comparison to the more recently formed C1 feature (Tab.~\ref{tab:dustem}~\&~\ref{tab:C1}). Our observations and analysis \rev{present} evidence of nanodust formation in the circumstellar environment around a dust-forming WC binary.

We discussed the possible mechanisms that could reduce the abundance of larger grains while enhancing the abundance of nanodust in the circumstellar environment of WR~140. The two most likely processes were shattering via grain-grain collisions and rotational disruption from radiative torques.
An analysis of the grain-grain collision and RATD timescales (Eq.~\ref{eq:tgg}~\&~\ref{eq:tratd}) suggests that both mechanisms are plausible in the circumstellar environment around WR~140. However, we slightly favor the RATD mechanism over grain-grain collisions as the dominant grain disruption mechanism due to the strong and hard radiation field from the WR~140's WR and O stars.

Following analytical formulas derived by \citet{Hoang2019} on the RATD mechanism, we showed that the strong radiation field of the central WC binary in WR~140 should be capable of disrupting grains larger than $a\gtrsim100$ \AA~at a distance of $\sim6000$ au from the central binary (Fig.~\ref{fig:RATD}). One of the major issues presented by the RATD mechanism, however, is that the short RATD disruption timescale of $\sim$days (Eq.~\ref{eq:tratd}) at the position of C1 suggests that the larger grain population should be rapidly and entirely disrupted, which disagrees with the presence of larger grains in the C1 Fossil. Further investigation of RATD and grain-grain collisions in the circumstellar environment of dust-forming WC binaries will be important for addressing the dominant dust-disruption / nanodust-production mechanism(s) from these Carbon-rich dust factories.

Lastly, we introduced the astrophysical implications of nanodust formation around dust-forming colliding-wind WC binaries like WR~140. We speculate that all dust-forming WC binaries should exhibit some degree of nanodust formation from the RATD mechanism given the strong radiation field emitted by the central WC star. Given the rapid evolutionary timescales to form WR stars ($\sim$Myr), WC binaries may therefore provide an early source of C-rich nanodust in the ISM of galaxies in the local and early Universe. As demonstrated in this work, high contrast and high spatial resolution IR observations will be crucial for investigating the impact of nanodust formation from WC binaries \rev{especially in the era of JWST}.

\acknowledgments
RML thanks Thiem Hoang for the insightful discussions on grain disruption and the RATD mechanism.
RML also thanks T.~Fujiyoshi, the Subaru Observatory staff, and the Keck Observatory staff for supporting our observations of WR~140. 
\rev{We also thank the anonymous referee for their valuable feedback on our analysis an interpretation.}
\rev{The work of RML is supported by NOIRLab, which is managed by the Association of Universities for Research in Astronomy (AURA) under a cooperative agreement with the National Science Foundation.}
RML acknowledges the Japan Aerospace Exploration Agency's International Top Young Fellowship (ITYF).
AACS is funded by the Deutsche Forschungsgemeinschaft (DFG, German Research Foundation) in the form of an Emmy Noether Research Group (grant number SA4064/1-1, PI Sander) and acknowledges additional support by the Deutsche Forschungsgemeinschaft (DFG, German Research Foundation) -- Project-ID 138713538 -- SFB 881 (``The Milky Way System'', subproject P04).
\rev{AFJM is grateful for financial aid from NSERC (Canada).}
\rev{IE acknowledges support from Grant-in-Aid for JSPS Fellows (grant No. 21J13200)}

This research is based on data collected at Subaru Telescope, which is operated by the National Astronomical Observatory of Japan.

The development of SCExAO was supported by the Japan Society for the Promotion of Science (Grant-in-Aid for Research \#23340051, \#26220704, \#23103002, \#19H00703 \& \#19H00695), the Astrobiology Center of the National Institutes of Natural Sciences, Japan, the Mt Cuba Foundation and the director’s contingency fund at Subaru Telescope. 

\rev{Some of the data presented herein were obtained at the W. M. Keck Observatory, which is operated as a scientific partnership among the California Institute of Technology, the University of California and the National Aeronautics and Space Administration. The Observatory was made possible by the generous financial support of the W. M. Keck Foundation.}

The authors wish to recognize and acknowledge the very significant cultural role and reverence that the summit of Maunakea has always had within the indigenous Hawaiian community, and are most fortunate to have the opportunity to conduct observations from this mountain.

CHARIS was built at Princeton University under a Grant-in-Aid for Scientific Research on Innovative Areas from MEXT of the Japanese government (\#23103002)

This research made use of Astropy,\footnote{http://www.astropy.org} a community-developed core Python package for Astronomy \citep{Astropy2013, Astropy2018}.

%

\vspace{5mm}
\facilities{Subaru/SCExAO(CHARIS), Subaru(COMICS), Keck II(NIRC2)}

\software{IRAF \citep{Tody1986,Tody1993}, COMICS Reduction Software, pyKLIP \citep{Wang2015}, CHARIS pipeline \citep{Brandt2017}, \dustem~ \citep{Compiegne2011}, Astropy \citep{Astropy2013,Astropy2018}}



\end{document}